\documentclass{appolb}
\usepackage{epsfig,amssymb,xspace}
\newcommand{\nc}{\newcommand}
\nc{\req}[1]{Eq.\,(\ref{#1})}  \nc{\eps}{\varepsilon}
\nc{\beq}{\begin{equation}}     \nc{\beql}[1]{\begin{equation}\label{#1}}
\nc{\eeq}{\end{equation}}       \nc{\rf}[1]{figure  \ref{#1}}
\nc{\beqa}{\begin{eqnarray}}   \nc{\eeqa}{\end{eqnarray}}       
     \nc{\pathlaptop}{/home/rafelski/figure/}
     \nc{\pathletes}{/users/lpthe/jletes/bookraf/figures/}
     \nc{\pathnow}{}
 \def\lessim{\lower.5ex\hbox{$\; \buildrel < \over \sim \;$}}
\def\gtrsim{\lower.5ex\hbox{$\; \buildrel > \over \sim \;$}}

\begin{document}
\title{Status of Strangeness-Flavor Signature of QGP%
\thanks{Lecture Notes for the Cracow School of Theoretical Physics, XLVIe Course, 2006.
}%
}
\author{Jan Rafelski
\address{Department of Physics, University of Arizona, Tucson, AZ85721}
\and
Jean Letessier
\address{Laboratoire de Physique 
                Th\'eorique et Hautes Energies\\
     Universit\'e Paris 7, 2 place Jussieu, F--75251 Cedex 05.}
}
\maketitle
\begin{abstract}
Is the new state of matter formed in relativistic heavy ion collisions the 
deconfined quark--gluon plasma? 
We survey the status of several strange hadron observables and 
discuss how these measurement help understand the dense hadronic matter.

\end{abstract}
\PACS{12.38.Mh,24.10.Pa,25.75.-q}

\section{Introduction}\label{Int}
Matter made of free quarks has been discussed in literature for 
more than  30 years. For further theoretical 
details and historical developments we refer 
to the  `Theoretical Foundations'
collection  of QGP articles~\cite{KMRbook}. 
Today, many believe
that a new  state of matter made of free quarks has been formed 
in relativistic nuclear collisions. Is this `really' 
 the   interacting plasma of QCD quanta, 
quarks and gluons (QGP), formed 
and present in a limited domain of space and  time, where
quarks and gluons are propagating  constrained by external 
`frozen  vacuum', which abhors color?
Before proceeding to  more specific matters,
we will  make a few general remarks. 

We shall
discuss in these notes the yields of a diversity of strange hadronic particles
produced in relativistic heavy ion collisions, and consider
also theoretical models of strangeness production in QGP.
In the analysis of (strange) hadron yields  we use  the 
statistical hadronization model (SHM) to connect
properties of matter with yields of particles (or vice-versa).
We note that  experiments detect reliably
just a small fraction of all hadrons of interest. 
In order to have  global and detailed information
about the properties of the dense matter source of
these hadrons,  an extrapolation  must be made to cover 
production yields of all particles. 

The yield of  measured particles are obtained from their spectra, which are 
extrapolated to allow the integration over the full phase space.   
This introduces considerable systematic uncertainty, but
reduces greatly the dependence on the dynamics of fireball 
evolution at time of hadronization. 
In this work, we address quantitatively only the $m_\bot$-integrated total, or 
rapidity density $dN/dy$, particle yields, noting sometimes 
the  spectral $m_\bot$ shape. Even in this reduced case, 
considering the complexity of  hadronization,
and  the very rich  experimental data set,
considerable effort is required to complete the data analysis.
For this reason,  we  helped  develop a hadronization tool
SHARE (Statistical HAdronization with REsonances),   
a  suite of  numeric analysis programs  which 
is  available in the public domain~\cite{share}. SHARE is
used in all analysis presented here. SHARE encompasses all
other  statistical hadronization models (SHM), which arise
as special cases in SHARE set of parameters.

As the above implies, 
there are several different approaches one can take
in the study of particle yields using SHM. In
our opinion, SHM is  a physics-motivated model 
extrapolating from a subset to  all particle yields. Therefore,
we take the most elaborate version
of the SHM model to compare with experiment: 
the results we present are obtained allowing for  the full
chemical non-equilibrium in the analysis of hadron 
yield data. 

We present here the  key strangeness QGP observables  and look at the
results with two questions in mind: has the experimental 
result been predicted to be a QGP consequence? Is an 
alternative explanation of the data available today, which does not
 contradict    the behavior of the data being explained?  
We will also discuss how understanding of strangeness production
at SPS and RHIC helps to prepare for the LHC energy range, and the 
low energy RHIC run.

We address in turn the   $\overline\Lambda/\bar p$ (next section),
Strange (anti)baryon  enhancement at SPS and RHIC (section \ref{ABEnh}),
The K$^+/\pi^+$-horn  (section  \ref{BMes}). In this context, we expand on
the observation of a change in the reaction mechanism, favoring baryons
at energies at or above the  K$^+/\pi^+$-peak. We then introduce
 the bulk observable, strangeness per entropy $s/S$ in section \ref{sSsec},
and show how it helps evaluate the ratio of degrees of freedom present.
We then show how  $s/S$ can be computed at RHIC and LHC and 
present results showing independence of the result from initial conditions
other than entropy content $dS/dy$. In section \ref{StrRes}, we discuss
strange hadron resonances and their importance to the diagnosis of QGP. 
  We close the discussion, in section \ref{Outlook},
 with a short outlook at the  developments in the future.

\section{ Ratio of $\overline\Lambda$ with  $\bar p$ at SPS}\label{aLapSec}
 The first proposed strangeness and
 hadron yield ratio signature  of QGP has been  the relative
yield  of $\overline\Lambda$ and $\bar p$~\cite{Rafelski:1980rk,abundance}.
After cancellation of combinatorial and phase space factors,
this ratio is  determined by relative quark yields
available at hadronization.  If no QGP were formed, 
one could at best hope for chemical equilibrium yields in 
the hadron gas matter, but especially at relatively low reaction
energies this requires a magic touch, the chemical equilibration 
of strangeness in hadron born reactions. 

For both particles considered we include, aside of directly produced 
$\overline\Lambda$, $\bar p$, the  
yields of more massive resonances  which decay into 
these particles.   SHARE  accounts
for this important effect. However  we present first   the more intuitive 
historical line of thought: we  assume that the resonance contributions  
multiply both $\overline\Lambda$ and $\bar p$ by  similar 
enhancement  factors.  This being the case,    resonance effect  can be 
ignored at first   when considering the $\overline\Lambda/\bar p$ ratio. 

In  a baryon-rich QGP environment, the 
light antiquark  $\bar u,\,\bar d$ abundances are suppressed.
This is easily understood as result
of  $\bar u,\,\bar d$ annihilation on $u,d$ excess present where baryons are
present. In the statistical hadronization model, this effect is described
by the  baryochemical potential $\mu_b=3(\mu_u+\mu_d)/2=3\mu_q$.
However,  the 
strange antiquark yield    $\bar s$ is also suppressed
by the strange quark mass. Integrating the particle phase space 
characterized by a production temperature $T_h$
 (hadronization temperature)  we find:
\begin{eqnarray}
 \left.\frac{\overline\Lambda}{\bar p}\right|_{\rm QGP}=
 \frac{N_{\bar s}N_{\bar u}N_{\bar d}}{N_{\bar u}N_{\bar u}N_{\bar d}}
& \simeq &\frac{\gamma_s^{\rm QGP}}{\gamma_q^{\rm QGP}}
   \frac 1 2 \frac{m^2_s}{T_h^2}K_2(m_s/T_h)
   e^{(\mu_u^{\rm QGP}-\mu_s^{\rm QGP})/T_h} \nonumber \\
&=&\frac{\gamma_s^{\rm QGP}}{\gamma_q^{\rm QGP}}\,0.9 \,e^{(\mu_u^{\rm QGP}-\mu_s^{\rm QGP})/T_h}.
\end{eqnarray}
The last equality follows for  the currently accepted value $m_s/T_h\simeq 0.7$.

The  relative  yield originating in
the hadron phase comprises, in place of strange quark mass 
suppression, the hadron phase space suppression factor:
\begin{eqnarray}
\left.\frac{\overline\Lambda}{\bar p}\right|_{\rm HG}&=&
 \frac{\gamma_s^{\rm HG}}{\gamma_q^{\rm HG}}
\left(\frac{m_{\overline\Lambda}}{m_{\bar p}}\right)^{3/2}
e^{-(m_{\overline\Lambda}\,-\,m_{\bar p})/T_f}
      e^{(\mu_u^{\rm HG}-\mu_s^{\rm HG})/T_f} \nonumber \\
&=& \frac{\gamma_s^{\rm HG}}{\gamma_q^{\rm HG}}
  \,1.3 \,e^{-180\,{\rm MeV}/T_f}e^{(\mu_u^{\rm HG}-\mu_s^{\rm HG})/T_f}
\end{eqnarray}
In the range $T_f\simeq 160\pm20$  MeV, the  HG yield is  
 significantly smaller  compared to the relative  yield from
 QGP. Moreover, in HG the multiplicative factor 
$\gamma_s/\gamma_q$,  will be well below unity at low
collision energy.  As a side remark, note that the analysis of the
experimental data suggests that this chemical equilibration 
factor  reaches unity for 
most central and most energetic SPS reactions which is 
evidence for a new production mechanism of all particles.

An important and self-evident result is that, using hadron 
phase space,  it is  
indeed impossible to ever obtain for $\overline\Lambda/\bar p$
a value  that exceeds unity. However,  the situation is  
different when  QGP hadronizes. Let us recall the equilibrium 
hadronization conditions:
\begin{equation} 
T_h \simeq T_f,\quad  
\langle s+\bar s \rangle^{\rm Q } \simeq \langle s+\bar s \rangle^{\rm H },\quad  
S^{\rm Q } \simeq S^{\rm H },\quad  
\mu_u^{\rm Q }-\mu_s^{\rm Q }=\mu_u^{\rm H }-\mu_s^{\rm H },
\end{equation}
here $S$ is the entropy and both $s$ and $S$ can be understood to be rapidity density. 
We abbreviate above  Q for QGP and H for HG. Moreover, irrespective of question
of phase transition, for  the purpose of  comparison of production 
yields  one normally assumes that   
chemical potentials are measured and thus yields are compared at a given 
value obtained from experimental data:
\begin{equation} 
\mu_u -\mu_s = {\rm Const.}\,.
\end{equation}
In either case much of  enhancement of $\overline\Lambda/\bar p$ (and other 
strange antibrayon relative yield  in QGP compared to HG) 
  is due to the reduced 
threshold of strangeness in QGP compared to HG.

This argument  correctly describes  the yields of dominant fractions of (strange)
particles which carry most of strangeness. For the rarely produced
particles which antibaryons are at low energy  the situation is   more involved. At the   onset 
of global hadronization,  $\mu_s^{\rm QGP}\simeq 0$. This is so since
 in QGP at first $s-\bar s=0$, and strangeness is not clustered in baryon
number objects in plasma.  Considering that  the  HG phase is 
asymmetric at a given baryon number,  there is a buildup of    $\mu_s^{\rm QGP} $
with time, reversing with time the asymmetric emission of hadrons. 

For conditions we consider, the  preferentially emitted particles are 
carriers of    $\bar s $-quarks. $\overline\Lambda$ (and $(q\bar s)$ kaons ) 
is emitted at first with preference.
As time progresses this leads to  an increase in $ \mu_s^{\rm HG}$ in plasma phase 
such that symmetric emission and later  $s$-quark emission dominance ensues, 
assuring  that strangeness
conservation $s-\bar s=0$  is achieved among all produced particles
by the end of hadronization. However, it has also been speculated that the early asymmetric 
emission of particles  with $\bar s>s$ could  lead  to the formation 
of a strangeness enriched residual matter~\cite{Greiner:1987tg}.
Extensive searches for this effect failed to confirm this reaction model. What is
perhaps instead happening
is that at some point $\mu_s^{\rm QGP} $ is large enough such that emission of 
particles  with $ s> \bar s $ dominates.

At the beginning of the process of hadronization, the production
of $\overline \Lambda$ (and $\overline\Xi,\overline\Omega$)
 is enhanced  by  $\mu_s^{\rm QGP}=0$. 
Once produced, particles do not disappear, and thus QGP property $\mu_s^{\rm QGP}=0$
is imprinted on the enhanced yield of rarely produced strange antibaryons, which are 
predominantly produced in the early time stage of QGP hadronization.  Thus, 
\begin{equation}  
 \left.\frac{\overline\Lambda}{\bar p}\right|_{\rm QGP}\simeq
0.9 \lambda_u^{\rm QGP},\qquad \lambda_u^{\rm QGP}=e^{(\mu_u^{\rm QGP} \!/T_h)}.
\end{equation} 
It is well understood that the value of quark fugacity 
 $\lambda_u^{\rm QGP}$ at hadronization 
increases with decreasing reaction energy. 
This implies  that with decreasing reaction 
energy the relative yield $\overline\Lambda/\bar p$ increases, 
which at first sight is an unexpected result. 

Beyond this  qualitative discussion, originally presented more than 25 years ago, 
a quantitative prediction which includes all resonance decays is shown in 
Fig. 47 of our Acta Physica Polonica 
review article from 1996~\cite{actab},
which is reproduced at hadronization with figure caption, see \rf{BARLP}.  
The energy scale  $E/B$  (fireball energy content per baryon) is somewhat
unusual, yet it corresponds almost exactly to $\sqrt{s_{\rm NN}}$,
the center of momentum frame nucleon pair reaction energy. Namely,
for stopping of energy being 50\% of participant stopping, 
the fireball energy content per baryon $E/B$, seen in \rf{BARLP}, would be 
just  the CM energy per colliding nucleon pair.  Study of energy and 
participant number content in the fireball matter formed at low energy 
SPS reactions suggest that 50\% relative stopping 
is the right order of magnitude.

\begin{figure}
\centerline{\psfig{file= \pathnow 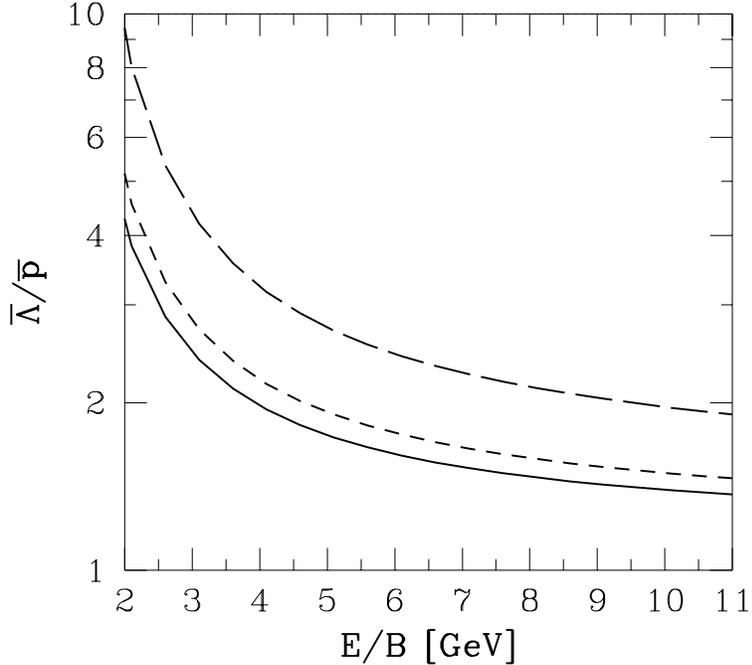,width=10cm}}
\caption{ 
Strange antibaryon ratio $\overline{\Lambda}/\overline{p}$,
as function of $E/B $ in a QGP-fireball for $\gamma_{\rm s}=1$;  
solid lines are for full phase space coverage, 
short dashed line  for particles with $p_\bot\ge 1$ GeV and
long dashed line for particles with $m_\bot \ge 1.7$ GeV (reproduced
from Acta Physic Polonica~\cite{actab}).
\label{BARLP} 
}
\vspace*{-3pt} 
\end{figure} 

\begin{figure}
\centerline{\psfig{file= \pathnow 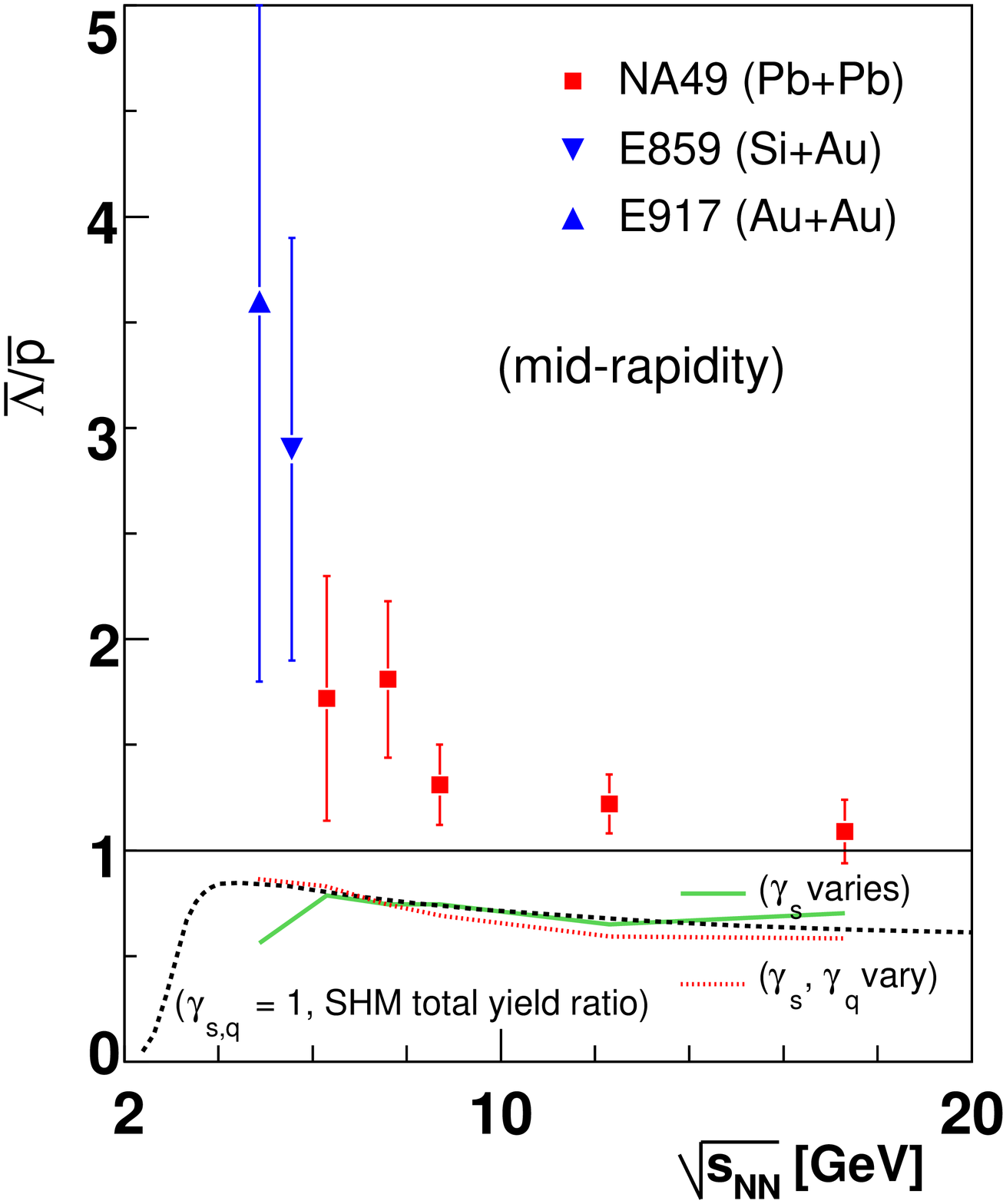,width=10cm}}
\vspace*{-13pt} 
\caption{
Top: observed mid-rapidity  particle yield  ratio 
$\overline \Lambda/\bar p$ as function 
of nucleon-nucleon reaction energy $\sqrt{s_{\rm NN}}$. 
Bottom: statistical QGP hadronization total yield ratios in different
QGP breakup scenarios.
NA49 compilation of own, AGS data and theoretical results.
\label{aLap}
}
\vspace*{-3pt} 
\end{figure}

The result seen in 
\rf{BARLP}  is   not surprising, following on our qualitative discussion, 
we see how the ratio   $\overline{\Lambda}/\bar{p}$ is increasing  
with decreasing reaction energy. It is of some interest
to compare this result with the  AGS and the SPS experimental  results.   
The data analysis has been evolving.
The results are shown in figure \ref{aLap}, based on compilation 
of  data and theoretical results by the NA49 collaboration~\cite{Alt:2006dk}. We  see
that  the central rapidity ratio  $\overline\Lambda/\bar p$ is well above unity at all available 
reaction energies. With decreasing reaction energy,  
this ratio  increases, just as it is expected in our above discussion. The
fact that to lowest considered collision energies and smallest reaction systems 
at AGS this trend persists (within the quite
large error bars) could suggest that QGP formation followed
by slow and   continuous hadronization is the reaction mechanism
governing the production of strange antibaryons  at relatively low
AGS energies.   

In the bottom of  figure \ref{aLap},  the bulk matter
predictions are shown, based on global SHM fits to the experimental data
carried out by different groups (including our own  results), under differing scenarios, 
all theoretical results have been assembled   by the NA49  group and we refer 
to this work for further discussion and references. Comparing  data  with
bulk yields  we see that the additional 
enhancement of $\overline\Lambda/\bar p$   we attribute to dynamically 
evolving $\mu_s$ is clearly most pronounced at lowest reaction energies.
In fact, for the  highest  reaction energies, where sudden breakup of the QGP fireball is
assured, the discrepancy
between bulk hadronization and experimental data could  perhaps be  accounted
for by conventional means. Specifically, the experimental  $\overline\Lambda/\bar p$   ratio needs to be reduced 
to correct for the included weak decays $\overline\Xi\to \overline\Lambda$ which are not included in the SHM models. 
Moreover,  the thermal rapidity distribution of  $\overline\Lambda$ is narrower than that
of $\bar p$, which enhances the central rapidity ratio compared to the $4\pi$-yields 
shown in the theoretical part of the figure.  On the other hand, the trend of the data at low energies is 
quite clear, despite  the large error bars.   

We conclude: $\overline\Lambda/\bar p>1$  ratio,  rising with decreasing reaction
energy, has been  predicted to arise in hadronization of QGP. No
other  explanation of this behavior is known to us. An alternate explanation would have to
address both the magnitude of the effect and its energy dependence.  The available  
  data is   consistent with QGP; however,  good experimental
results are needed to confirm   this intriguing trend. 

On the other hand, we   note
that the absolute yields $\overline\Lambda$, and $\bar p$ are not large. 
At the top AGS energy (11.6 $A$ GeV for central Au--Au collisions),
in the most central reactions, one 
$\overline\Lambda$ or $\bar p$ will be produced in one out of 100 Au--Au collisions. 
There are important kinetic rescattering 
 processes which can generate a shift $\overline\Lambda\leftrightarrow \bar p$  
inducing  suppression, or enhancement of  $\overline\Lambda$ or $\bar p$.
Qualitative kinetic model rescattering arguments favor  $\overline\Lambda$ over $\bar p$
yield: the annihilation cross section on baryon matter is smaller for   $\overline\Lambda$.
The strangeness exchange cross reaction is exothermic for 
$\bar p+(q\bar s)\to \overline\Lambda +(q\bar q)$. Thus, kinetic rescattering 
in HG  could   increase the ratio $\overline\Lambda/\bar p$ beyond 
relative  chemical equilibrium yield.

Such dynamical effects are not common but 
must be investigated before we conclude that QGP is formed at  top  AGS energies
When good data at AGS reaction energy becomes available,  $\overline\Lambda/\bar p$   is one 
of the observables meriting very careful experimental and theoretical study.  This is the low reaction energy
probe of QGP most resembling $J\!/\!\Psi$ suppression: the signal is very clean, but must be
carefully considered for competing effects. 

 \section{Strange (anti)baryon  enhancement at SPS and RHIC}\label{ABEnh}
CERN experiments WA97 and NA57 have focused  on the study 
of the  systematics  of the strange (anti)baryon enhancement 
with reaction energy, and centrality in Pb-Pb collisions.  $\Lambda, \Xi$ and $\Omega$ and antiparticle 
yields have been  measured at central rapidity $y$
and medium transverse momentum $p_\bot$   as functions of the centrality of the collision. 
Comparing the yields in Pb-Pb to those in $p$Be interactions, considerable enhancement  of yield per 
participant is   observed. This  enhancement increases linearly with the centrality, and geometrically
with the strangeness  content in hyperons, reaching a factor of about 20 
for the $\Omega+\overline\Omega$ in the central Pb-Pb collisions. 
Recently, the final results for the centrality 
dependence of  (anti)hyperon production in Pb-Pb, $p$Pb and $p$Be collisions at 158 $A$ GeV/c  
have been reported~\cite{Antinori:2006ij}.  The open symbols, in  \rf{ABEnhance}, show these
results, which  follow the pattern predicted in the recombination-hadronization 
model~\cite{Koch:1986ud}.  
The enhancement rises, both with the strangeness content in  the hadron, and 
with the participant number $A$ (centrality), that is the size of the reaction region.
The magnitude of the the enhancement is nearly the same as seen at   much lower SPS energy range.

\begin{figure}
\centerline{
\psfig{width=11cm, figure=\pathnow  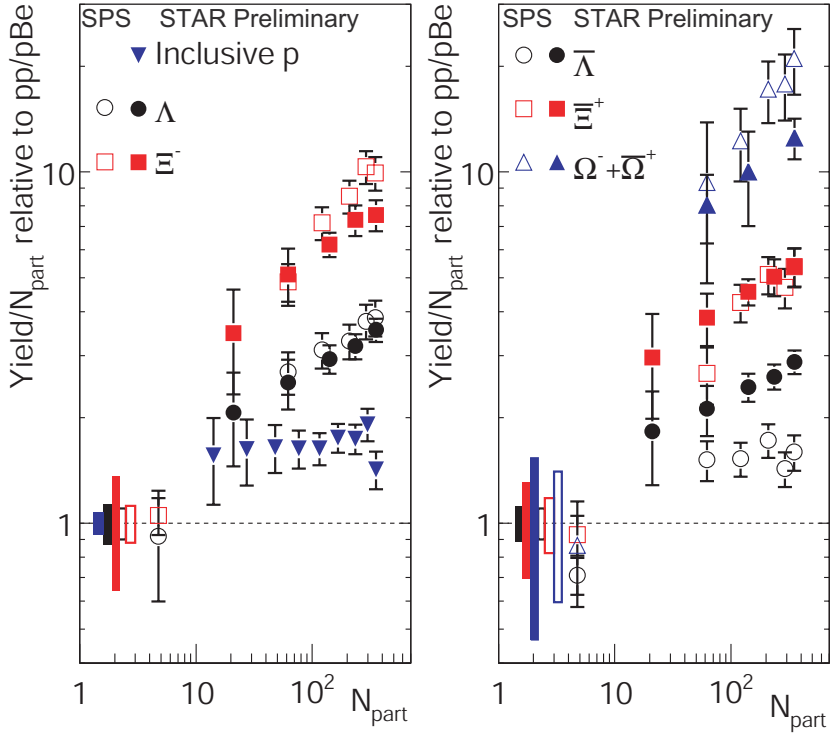}
}
\caption{
\label{ABEnhance} 
Yields  per participant $N_{\rm part}$, for NA57-SPS  Pb--Pb $\sqrt{s_{\rm NN}}=17.3$\,GeV  (open symbols) relative to 
$p$Be  and for STAR-RHIC Au-Au $\sqrt{s_{\rm NN}}=200$\,GeV  
(filled symbols) relative to $pp$. On left baryons, on right 
antibaryons and $\Omega+\overline\Omega$ (triangles), circles are   
 $\Lambda$  and  $\bar{\Lambda}$,  squares are $\Xi^{-}$ and $\bar{\Xi}^{+}$.
Error bars represent those from the heavy-ion measurement. 
Ranges for $pp$ and $p$Be reference data   indicate the statistical and systematic
uncertainty. After~\cite{Caines:2006if}.  
}
\end{figure}
 
There are   further features of the  WA97 and NA57 experimental results which 
are particularly important: the enhancement seen at 40 $A$ GeV/c is very similar, 
nearly identical, particle by particle, as  seen 
at 158 $A$ GeV/c~\cite{Antinori:2004ee,Antinori:2006rk}  expert for the most
peripheral class of events. We read this   result  to mean 
 that  the peripheral 40 $A$ GeV/c  reactions do not always reach deconfinement. 
The final experimental results  differ little from the preliminary conference
reports which we discussed  at this school  in   2003~\cite{Rafelski:2003ju} 
and  in 1999~\cite{Rafelski:1999xu}. 
For this reason, we focus discussion of this enhancement pattern  at RHIC, 
and the comparison of RHIC  with SPS. 

The solid symbols, in  \rf{ABEnhance}, correspond
to the most recent STAR results obtained 
at $\sqrt{s_{\rm NN}}=200$ GeV~\cite{Caines:2006if}.
The yield of (multi)strange baryons  $\Lambda(uds)$, $\Xi^-(dss)$
and antibaryons per  participant   $N_{\rm part}$ in the reaction,  divided 
by a reference yield obtained in $pp$ reactions is shown. 
Within error, considering also the base yield, the enhancement at
RHIC and SPS appears the same. However, there are two issues to consider:\\
a) The enhancement computed by NA57 is based on   $p$Be, where
 some enhancement of $\Lambda$  is expected to be present, 
as compared to $pp$ results.  The NA49 SPS
experiment has just reported $\Lambda$ enhancement  
evaluated with reference to $pp$~\cite{Mitrovski:2006js}.
In this more directly comparable case,  the SPS $\Lambda$-enhancement comes to be
factor 5-6 (rather than 4-5) in most central Pb-Pb reactions.\\
b) The behavior of   $\overline\Lambda$ enhancement at SPS
 breaks the ranks   in that it is seen to be nearly flat as function
of centrality, and it is smaller at SPS than at RHIC. \\
 We conclude that the enhancement pattern of $\Lambda$ and $\overline\Lambda$ is 
influenced by the prevailing baryon density,   for $\overline\Lambda$ the enhancement 
is greater at RHIC where the baryon density is smaller than at SPS, while the
reverse is true for $\Lambda$. This baryochemical potential effect is mostly
erased considering d $\Xi^{-}$, $\overline{\Xi}^{+}$, but is visible within the error bar. 
The enhancement of  $\Omega+\overline\Omega$ is largest, since production of
these particles is very difficult in the elementary reactions, especially so at the 
lower SPS energy. 

\begin{figure}
\centerline{
\psfig{width=12cm, figure=\pathnow  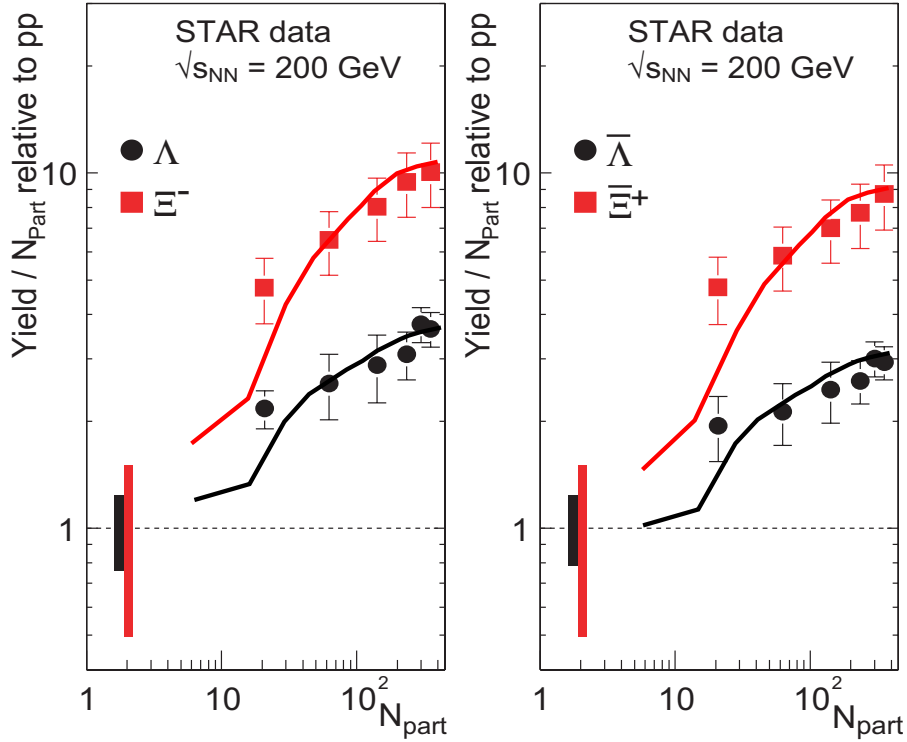}
}
\caption{
\label{RHICEnhance} 
Yields  per participant $N_{\rm part}$ relative to $pp$  of 
 $\Lambda$ and $\Xi^{-}$ on left  and  $\overline{\Lambda}$ and $\overline{\Xi}^{+}$
on right   in Au+Au collisions 
at $\sqrt{s_{\rm NN}}=200$\,GeV~\cite{Caines:2004ej} (compare \rf{ABEnhance}).
 Our theoretical results~\cite{Letessier:2005kc}
are for  chemical  non-equilibrium  and are normalized to the same $pp$ experimental 
results.
}
\end{figure}

The interpretation we pursue is that  the  strange antibaryon enhancement   
is due to   an increased yield density of strange quarks at hadronization,
growing within the   geometric source size, the increase driven by the longer  lifespan.  The gradual 
increase of the enhancement over the range of $N_{\rm part}$ is an important
indicator of the kinetic strangeness production mechanism. We return to this  
 in section~\ref{Ssec}, in  the   strangeness yield analysis 
as function of  centrality~\cite{Letessier:2005kc}. This enhancement at RHIC 
is in fact fully understood.   We compare the theoretical predictions 
with the experimental results for 
 $\Lambda$ and $\Xi^{-}$ on left  and  $\overline{\Lambda}$ and $\overline{\Xi}^{+}$
at RHIC in  figure~\ref{RHICEnhance}. Note that the normalization of the 
enhancement is different~\cite{Caines:2004ej}, as compared to \rf{ABEnhance}.  
This has occurred since the interpretation 
of the reference  $pp$ data changed. However, this vertical shift 
(in a log-figure) 
does not affect the comparison of theory and experiment since 
both use the same experimental $pp$ yield normalizer to define the
enhancement. We note that data in figure \ref{ABEnhance} is ``newer''.

The theoretical  lines, in  figure \ref{RHICEnhance}, follow from study of strangeness
yield in these reactions based mainly on PHENIX 
single strange hadron results~\cite{phenixyield},  and 
application of statistical hadronization model using these results~\cite{Rafelski:2004dp},  
 to predict the  yields shown~\cite{Letessier:2005kc}.  For most peripheral collisions, 
this study is not entirely applicable, since the SHM fit to the data is not fully reliable. This
is the `kink' in the theoretical curves seen in \rf{RHICEnhance}. An important help in understanding
these results is the gradual growth of the enhancement with centrality. As we see, the trend is
quantitatively consistent with strangeness yield growth (see section \ref{Ssec}). Both the 
energy and centrality dependence of the enhancement contradicts the canonical suppression
model, which reinterprets this enhancement as being due to reference 
data suppression~\cite{Redlich:2001kb}.

\section{From Baryons to Mesons}\label{BMes}
\begin{figure}
 \centerline{
\psfig{width=10cm,clip=,figure=\pathnow 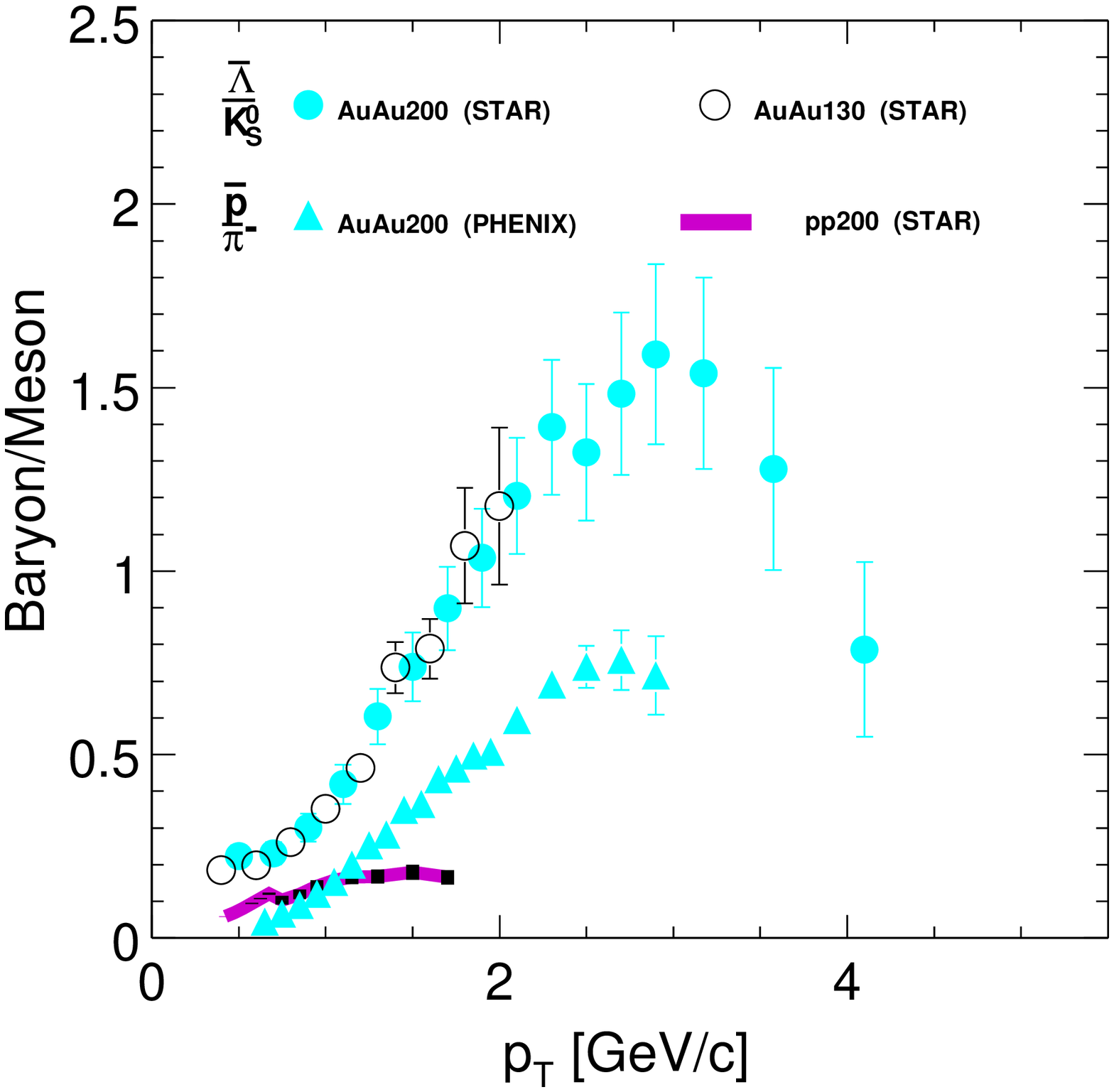}
}\vspace*{-0.3cm}
\caption{Ratios of ${\overline{\Lambda}}$ to $K_{S}$ from Au-Au and $pp$ collisions (STAR) and 
$\overline{p}$ to $\pi$ from Au-Au collisions (PHENIX) as a function of transverse momentum ($p_\bot$).
In addition to resonance contributions in all hadrons, the $\overline{\Lambda}$  includes 
contributions from $\Sigma^{0}$ decays.}
\label{ratio}
\end{figure}

The enhancement of (strange) (anti)baryons we
just presented is in part   due  
 to enhanced production of baryons as compared to mesons when 
AA reactions are compared to $pp$ reactions.
The figure~\ref{ratio} shows the ratios of
 $\overline{p}/\pi $ and $\overline{\Lambda }/K_{S}$ from central Au-Au collisions at 
$\sqrt{s_{NN}}=130,$ and 200 GeV measured by PHENIX~\cite{phenixyield} 
and STAR~\cite{star-b1,star-b2,star-k}.

We  recognize that the formation of mesons and baryons occurs
by a different mechanism than in the elementary $pp$ reactions.
In particular, the large baryon to meson ratio, seen in figure~\ref{ratio},  
cannot be accommodated by the conventional string  fragmentation scheme 
developed for the   elementary $e^{+}e^{-}$ and $pp$
reactions. The quark recombination models~\cite{Duke,Hwa} 
provided a satisfactory description of
the particle yields, in particular, the large production of anti baryons in the
intermediate $p_\bot$ region.
 
Within the statistical hadronization model, when chemical equilibrium is assumed,
the relative ratio of baryons to mesons is fixed, and varies along with  hadronization
temperature $T$.  When chemical nonequilibrium ideas are introduced,
the parameter controlling the 
relative abundance of baryons with respect to mesons is $\gamma_q$,
\begin{equation}
\frac{\rm Baryons}{\rm Mesons}=\gamma_q R_P(T,\frac{\gamma_s}{\gamma_q}, \frac{\mu_i}{T}),
\end{equation}
where $\mu_i$ are  chemical potentials, and 
$ \gamma_s/\gamma_q$ is the relative strangeness to light quark phase space occupancy. 
There is excess baryon-antibaryon pair yield over 
chemical equilibrium for $\gamma_q>1$,  expected
when the source of particles is a very dense deconfined quark state.   
Conversely, for low energies or small systems, we expect $\gamma_q<1$.
The value $\gamma_q=1$ can be established when there is time to 
scatter and equilibrate the yields of mesons and baryons, which in
general is not the case in all reactions described here. 

\begin{figure}
 \centerline{
\psfig{width=10cm, figure=\pathnow  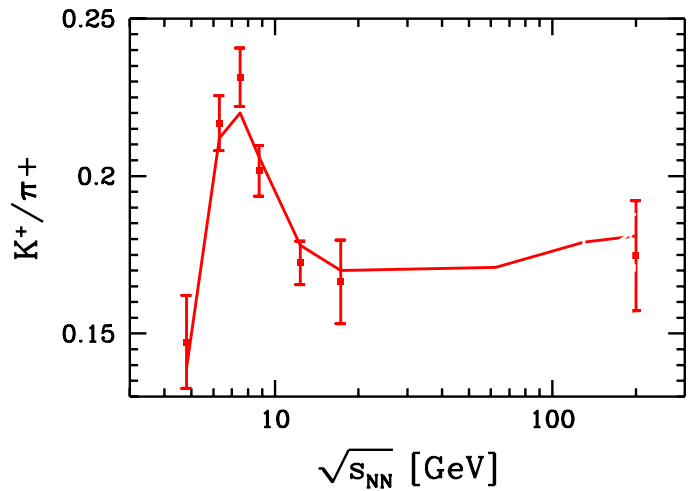}
}\vspace*{-0.3cm}
\caption{${\rm K}^+/{\pi}^+$ ratio of NA49 experiment~\cite{Gaz,Lee03}. Solid line: Result of 
chemical nonequilibrium hadron production analysis~\cite{Letessier:2005qe}.}
\label{kpiratio}
\end{figure}

Consideration of chemical nonequilibrium and the possibility that in some
reactions  $\gamma_q>1$, and in others  $\gamma_q<1$, provides an
opportunity to  describe the energy dependence of particle production,
studied by the experiment  NA49 ~\cite{Letessier:2005qe}. This has as special objective
the  interpretation of the   ${\rm K}^+/{\pi}^+$ ratio~\cite{Gaz,Lee03}, which   shows a 
pronounced energy dependent structure, see figure~\ref{kpiratio}.

The rapid rise in   the ${\rm K}^+/{\pi}^+\propto \bar s u/\bar d u$ ratio is due to 
the more  rapid increase in $\bar s$ than $\bar d $ when the baryon density 
is very high.   The decrease beyond the peak is driven by a
reduction of baryon stopping with   increasing   reaction energy, at which time
there is  a faster increase in   the $\bar d$ yield. 
In the  SHM  description the key ingredient  needed to obtain 
 this behavior, in consistency with all particle yields,
is the switch-over  at the peak of the ratio, 
from $\gamma_q<1$ to  $\gamma_q>1$. The gradual rise of  ${\rm K}^+/{\pi}^+$
at high energy is associated with an increase of $\gamma_s/\gamma_q$. 

The observable  ${\rm K}^+/{\pi}^+$ has been long considered as a  signature of 
strangeness enhancement~\cite{Glendenning:1984ta}. It turned out to also   be
a signature of anti-quark suppression in high baryon density fireballs created 
at low reaction energy. Moreover, there appears to be a change in hadronization
mechanism associated with the intermediate peak in  ${\rm K}^+/{\pi}^+$.
The advancement of this most interesting physics result is mainly due to the 
diligent work of Marek Ga\'zdzicki~\cite{Gaz}.

\section{Strangeness per Entropy} \label{sSsec}
Individual particle yields are in general probing conditions prevailing at the 
time of QGP  hadronization. However,   bulk  fireball  properties  
 reach more deeply into the history of QGP evolution. 
Of particular interest is 
the ratio strangeness per entropy $s/S$.  
Studies of QGP   suggest that entropy and 
to a lesser extent strangeness yields are produced in the 
earliest stages of the heavy ion reaction.  In the next section \ref{Ssec},
we will show how `deep' strangeness probes. In this section, 
we look at the qualitative features of this observable. 

The deconfined state
 is rich in both in entropy and strangeness. 
The enhancement of entropy $S$ arises
because the color bonds are broken and gluons can be created. 
Enhancement of  strangeness $s$ arises  because the 
mass threshold for strangeness excitation is considerably lower 
in QGP than in hadron matter. While entropy $S$ determines the hadron multiplicity content,
strangeness $s$ determines strange hadron content.  Thus, to
measure $s/S$, we need to have measured particle yields which  
comprise much of the total strange and non-strange hadron multiplicity.

 In QGP, strangeness $s$ originates primarily in gluon based reactions,
\eg, $GG\to s\bar s$, and higher order gluon related  processes. Gluons are the 
source of strangeness;  their presence requires deconfinement, 
and  is a result of entropy production, \ie, 
excess entropy $S$ content. It is generally 
believed that the entropy production is completed 
after first 0.5--1 fm/c reaction time, at which time the parton 
matter has turned into the thermalized QGP phase. 
Strangeness is   produced shortly after. We have been able to 
unravel much of mystery behind strangeness production (see section \ref{Ssec}). 
However, mechanisms of entropy formation are not  understood.

We estimate the expected magnitude of $s/S$.  For thermally  equilibrated 
non-interacting QGP phase at a  temperature $T_s$ at which formation of 
strangeness essentially has ended:
\beql{sdivS}
{s \over S}\equiv\frac{\rho_{\rm s}}{\sigma}     =
\frac{ (3/\pi^2) T_s^3 (m_{  s}/T_s)^2K_2(m_{  s}/T_s)}
  {(32\pi^2/ 45)  T_s^3 
    +n_{\rm f}[(7\pi^2/ 15) T_s^3 + \mu_q^2T_s]}=
{0.03\over {1+ 0.054 (\ln \lambda_q)^2} }\,.
\eeq
Here, we used for the number of flavors $n_{\rm f}=2.5$ and $2m_{  s}/T_s=1$. We 
see that the result is a slowly changing function  of $\lambda_q$;
for large $\lambda_q\simeq 4$, present at modest SPS energies, the 
value of $s/S$ is reduced by 10\% as compared to  $\lambda_q\simeq 1$ applicable to LHC and RHIC. 
Considering 
the slow dependence on $x=m_{  s}/T\simeq 1$ of $W(x)=x^2 K_2(x)$ there is 
a minor residual dependence on the value of $T_s$.

The dependence on the degree of chemical equilibration 
which dominates is easily obtained by separating the different 
degrees of freedom:
\beql{sdivS2}
{s \over S}=0.03 {\gamma_s^{\rm QGP} \over 
  {0.38 \gamma_{\rm G}^{\rm QGP}+ 
         0.12 \gamma_s^{\rm QGP}+
         0.5\gamma_q^{\rm QGP} + 
         0.054 \gamma_q^{\rm QGP} (\ln \lambda_q)^2}}\,.
\eeq
Given  \req{sdivS2}, we expect to see a gradual 
increase in $s/S$ as the QGP source of 
particles approaches chemical equilibrium with increasing 
collision energy   and/or  increasing volume.

How does this simple prediction compare to experiment?
A precise measurement of $s/S$ is not easy,
since it requires measurement of several particles
species to considerable precision.  Analysis of 
existent data was done as  a function of centrality~\cite{Rafelski:2004dp}, and  
as a function of reaction energy~ \cite{Letessier:2005qe}. 
The centrality study included 11 centrality bins of Au-Au at  $\sqrt{s_{NN}}=200$\,GeV
for which $dN/dy$ for   $\pi^\pm, {\rm K}^\pm, p$ and $\bar p$ at 
$y_{\rm CM}=0$ have been  
presented in table VIII, in Ref. \cite{phenixyield} (PHENIX). 
These 6 particle yields and their ratios change rapidly as function of centrality.
In addition results from     STAR  for 
${\rm K}^*(892)/{\rm K}^-$ \cite{haibin2200},
and  $\phi/{\rm K}^-$  \cite{phiyld}  which show 
minor centrality dependence were used.
  
Supplemental constraints
(ratio   $\pi^+/\pi^-=1.\pm0.02$,  strangeness conservation, electrical charge to net baryon ratio)
help to determine the best fit.
The  7 SHM parameters (volume per unit of rapidity 
 $dV/dy$, temperature $T$, four chemical 
parameters $\lambda_q, \lambda_s, \gamma_q, \gamma_s$ and the 
isospin factor $\lambda_{I3}$ are in this case
studied in  a systematic fashion as function of impact parameter using 
11 yields and/or ratios and/or constraints, containing one (pion ratio) 
redundancy~\cite{Rafelski:2004dp}.

  $\chi^2$ minimization  yielding good
significance is easily accomplished, showing good consistency of the
data sample. The resulting statistical parameters are seen
in \rf{gammu}, on left, as function of participant number. 
The full  non-equilibrium model  $\gamma_q\ne 1, \gamma_s\ne 1$ (full circles, blue) 
and semi-equilibrium  $ \gamma_q= 1$ (open circles, red) are shown. From top to bottom the  (chemical)
freeze-out temperature $T$, the occupancy factors 
$\gamma_q$, $\gamma_s/\gamma_q$ and together in the
bottom panel  the baryon   $ \mu_B$ and hyperon $\mu_S$ 
chemical potentials. 

\begin{figure}
\centerline{\hspace*{-.1cm}
 \psfig{width=6.5cm,figure=\pathnow 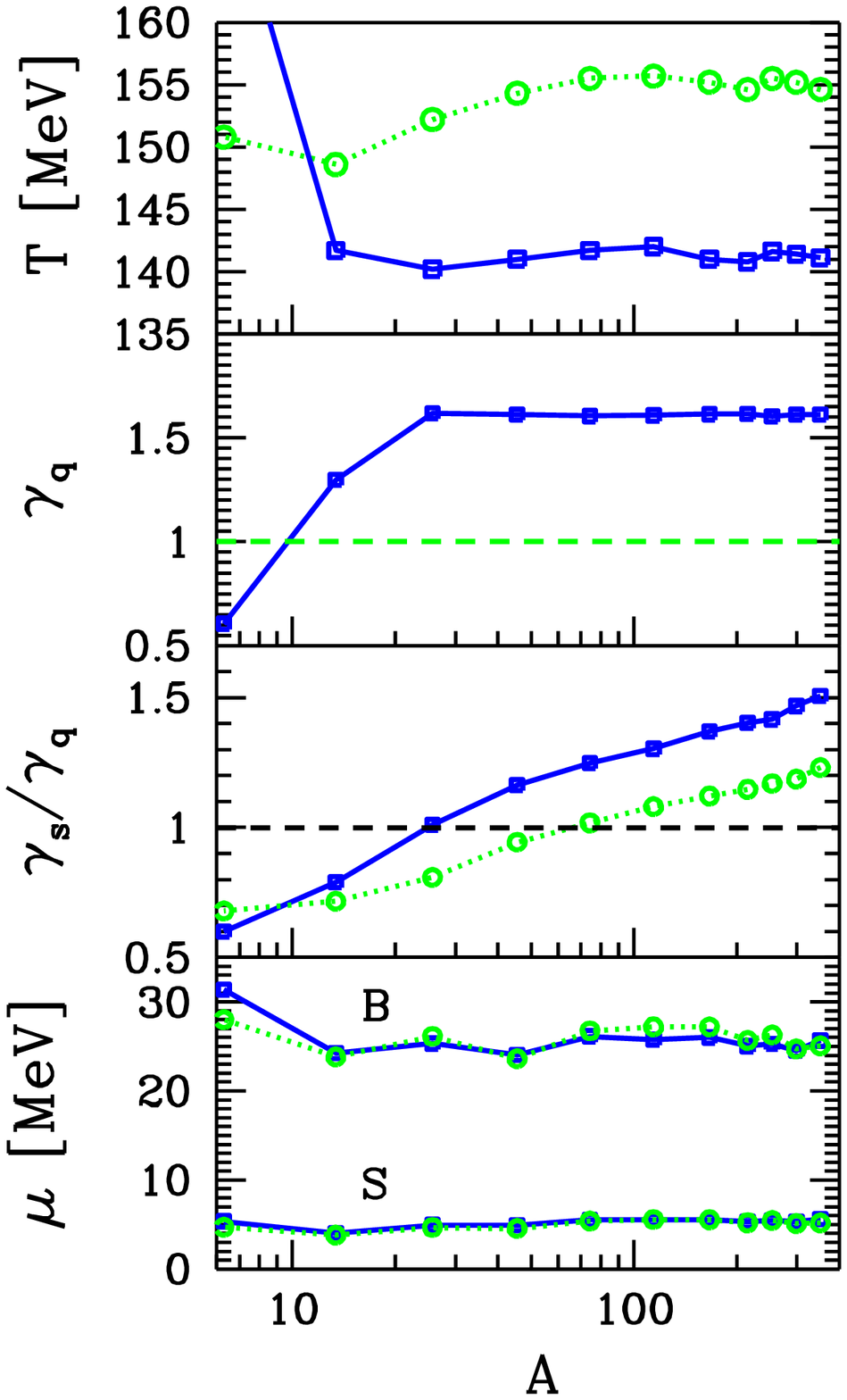} 
 \hspace*{-.50cm}\psfig{width=6.5cm,figure=\pathnow    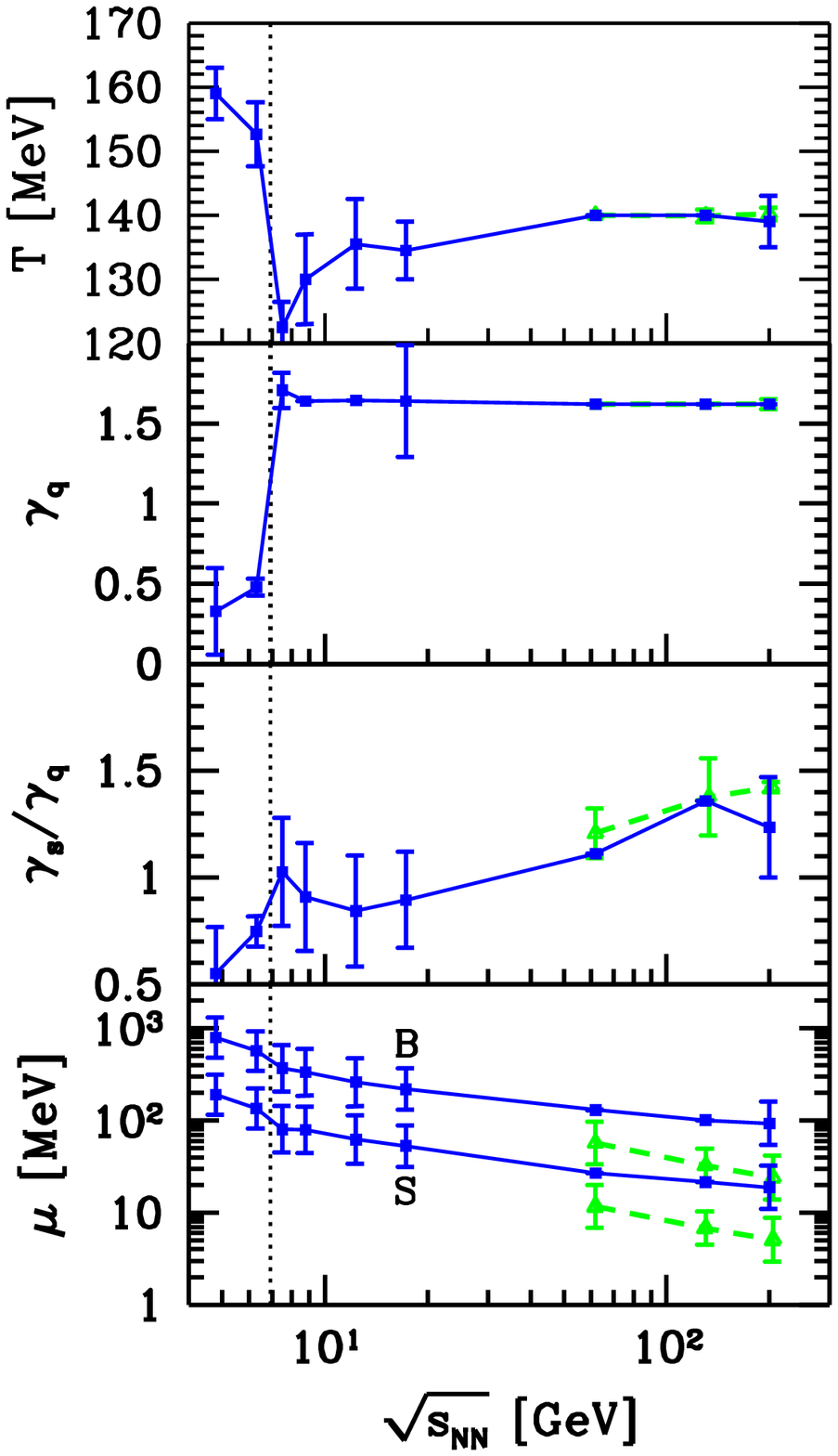}
}
 \caption{ \label{gammu} 
 From top to bottom: temperature $T$, light quark phase 
space occupancy $\gamma_q$,
the ratio  of strange to light quark phase 
space occupancies $\gamma_s/\gamma_q$  and the chemical potentials
($B$ for baryochemical $\mu_B$ and $S$ for strangeness  $\mu_S$) 
The lines guide the eye.
Left:  
as function of inelastic (wounded) participant
number $A$. The full  non-equilibrium model  $\gamma_q\ne 1, \gamma_s\ne 1$ (full circles, blue) 
and semi-equilibrium  $ \gamma_q= 1$ (open circles, red) are shown. 
Right: 
energy $\sqrt{s_{\rm NN}}$-dependence in full chemical nonequilibrium analysis.
 Note that the additional 
results at RHIC (two highest energy results  in green on right) 
 apply to central rapidity region $dN/dy$ study, while the AGS/SPS 
and Brahms-RHIC results   are for  total  yields (blue lines).
}
\end{figure}

Each of the panels shows very interesting behavior indicating 
some particular physics feature at chemical freeze-out. 
Perhaps the overall most striking result is that, 
 for $A>20$ only, $\gamma_s/\gamma_q$ 
varies significantly, as does  (not shown here) the volume of the reaction 
domain $dN/dV$. The scaling of $dN/dV$
with $A^2/3 $~\cite{Letessier:2005kc} is expected in a
geometric reaction picture. 
The growth of $\gamma_s/\gamma_q$
indicates that with size of the system there is more 
effective cooking of strangeness.  
There is no saturation of $\gamma_s$ as we approach the most
central reactions. This is inherent in the data we consider
which includes the yields of $\phi$, and $K^*$. This result is consistent
with the implication that strangeness is not fully saturated in the
QGP source, though it appears over-saturated when measured in 
the hadron phase space.

The baryo-chemical potential and hyperon potential (bottom
panel) is model independent and its nearly constant 
value indicates that only the reaction energy and not the centrality
of the reaction establishes baryon density at  hadronization.  
We find a hadronization
temperature  which, when we allow for $\gamma_q\ne 1$, is
at the level of $T=140$ MeV, and it rises to $T\simeq 155$
when we force light quark chemical equilibrium on the 
particle yield fit. 

On the right, in \rf{gammu}, we show the same statistical
variables as a function of reaction energy.
To study the energy dependence,  we   must assemble  several
different experimental results from different facilities and 
experiments~\cite{Letessier:2005qe}.  The lowest energy result is from our 
AGS study~\cite{Letessier:2004cs}; the SPS data we used are from NA49 energy dependence 
exploration at the CERN-SPS~\cite{Gaz}. These results are
for the total particle yields. The two highest energy  results are based on 
studies of RHIC data at 130 and 200 GeV  at central rapidity and address the $dN/dy$
particle yields, with the highest point corresponding to the 
results presented at greatest centrality on the left-hand side in \rf{gammu}.

There are several energy dependent   
 features seen on the right-hand side of \rf{gammu}.  We note 
the dotted vertical line  at energy where  some of the SHM parameters 
undergo a change of behavior. The value of energy corresponds
to where NA49 reports a change in the behavior of 
K$^+/\pi^+ \propto \bar s/\bar d$.

The statistical parameters obtained in the analysis 
are used to compute $s/S$ as
function of centrality~\cite{Rafelski:2004dp}, and  
as function of reaction energy~ \cite{Letessier:2005qe}. 
One indeed finds that corresponding to the growth of 
$\gamma_s/\gamma_q$ seen in \rf{gammu}, the 
value of $s/S$ grows, reaching nearly the 
QGP chemical equilibrium value, $s/S\simeq 0.03$
at highest RHIC energy in most central reactions.  
As a function of centrality at top RHIC energy    flat peripheral 
behavior where $s/S\lessim 0.02  $  at $A<15$ turns into 
smoothly increasing $s/S$
 reaching $s/S\simeq 0.03$ in most central reactions. 
As a function of energy near to $\sqrt{s_{\rm NN}}\simeq 6.5$ GeV
in most central collisions, there is a change in growth rate of $s/S$.

The significance of $s/S$ as an observable of QGP 
is that it is not dependent on the details of the dynamics of hadronization
which remain unresolved today. The value is predictable, and helps us 
understand the degree of chemical equilibration of QGP prior to hadronization.
The numeric value for saturated QGP phase is directly proportional to the 
ratio of the number of strange degrees of freedom to all degrees of freedom 
in the plasma ($s/S\propto g_s/4g$, see next section). Thus, $s/S\simeq 0.03 $ indicates
that   quark--gluon quanta are active. 
The smooth rise and the magnitude of $s/S$ are consistent with the 
QGP reaction picture --- a full interpretation of $s/S$ requires better 
understanding of QGP chemical equilibration, a topic we address next.

\section{Strangeness production at RHIC and LHC}\label{Ssec}
The kinetic description of  strangeness production can be reorganized 
to directly  evaluate the  strangeness yield normalized 
by entropy yield in plasma, $s(T(\tau )/S(T(\tau ))$. 
In this way, the model dependence on the
ex\-pansion-dilution  phenomenon  due to rapid expansion of
the QGP phase is considerably reduced.  We describe our recent
progress in the study of the chemical equilibration
at RHIC and LHC~\cite{Letessier:2006wn}, and evaluate 
how this work supports the interpretation of the results of 
data analysis for $s/S$ presented above.

The temporal evolution of $s/S$, in an expanding plasma, 
is governed by:  
\begin{eqnarray}\label{qprod3a}
 {d\over { d\tau}} {s\over S}
=
 {A^{gg\to s\bar s}\over (S/V) } 
    \left[\gamma_{\rm g}^2(\tau)-\gamma_{s}^2(\tau)\right]  
+
 {A^{q\bar q\to s\bar s}\over (S/V) } 
    \left[\gamma_q^2(\tau)-\gamma_{ s}^2(\tau)\right]\,.
\end{eqnarray}
When all $\gamma_i\to 1$, the Boltzmann collision term vanishes, the system has
reached chemical equilibrium. The value arrived at for the observable $s/S$ depends
on the history of how the system evolves at early stages,  and reaches QGP 
chemical equilibrium for  gluons in particular. 

In order to be able to solve  Eq.\,(\ref{qprod3a}),   a relation 
between $s/S$ and   $ \gamma_{ s}$   is required:
\begin{equation}\label{sS1}
\frac sS=\gamma_s  {g_s \over g} \frac{90}{8\pi^4}z^2K_2(z),\quad z=m_s/T,\quad
g_s=2_s3_c\left(1-\frac{ k\alpha_s(T)}{\pi}\ldots\right).
\end{equation}
 In the initial period,  gluons and quarks have not reached chemical equilibrium,
thus the actual
numerical integrals of Bose and Fermi distributions are
dependent on the values $\gamma_{q,{\rm g}}$.  The quark--gluon QCD  interactions
are introduced to describe
the effective degeneracy with precision:
\begin{equation}\label{ggq}
g=2_s8_c\gamma_{\rm g}\left(1-\frac{ 15\alpha_s(T)}{4\pi}\ldots\right)
 +\frac74 2_s3_c2_{\rm f} \gamma_{q}\left(1-\frac{ 50\alpha_s(T)}{21\pi}\ldots\right)+\,{\rm strangeness}.
\end{equation}
The strangeness component is made consistent with $g_s$ in Eq.\,(\ref{sS1}) and the concurrently 
 numerically computed $\gamma_s$. 

The degeneracies, seen  in Eq.\,(\ref{sS1}) 
for the entropy,  include the effect of interactions. Therefore, 
consistency requires that   the interaction effect is also introduced in the
strange quark degeneracy $g_s$, \req{sS1}.
The value of $k=2$ applies to massless strange quarks. At $T=0$ 
(or said differently, for $m\gg T$) the early study of quark matter 
self-energy  suggests that $k\to 0$~\cite{Chin:1979yb}.  The full range $0<k<2$ 
has been explored, the results for $k=1$ are presented.

We need to model the expansion dynamics of QGP matter. 
The study of hadronization at 
RHIC provides a strong constraint on the initial condition and 
 the evolution dynamics of QGP at RHIC. 
For LHC  three modifications are introduced:\\ 
\indent 1. To account 
for the greater reaction energy, we increase
 the initial entropy $dS(\tau_0)/dy$  by factor 4 from 5000 at RHIC to 20,000 at LHC. We do 
not   change in the initial value of $s/S|_{\tau_0}=0.016$; thus, 
in elementary parton-string  interactions, the relative strength of strangeness
and non-strange hadron production is left unchanged. This implies 
  an increase in initial strangeness yield by a
factor 4 at LHC compared to RHIC.\\
\indent 2.
The initial parton  thermalization time is reduced 
from $\tau_0=1/4$ fm at RHIC to  $\tau_0=1/10$ fm at LHC. This does not 
influence the outcome. \\
\indent 3.
In order to accommodate the greater transverse expansion pressure, 
we increased the maximum transverse
flow velocity reached to  $v_\bot=0.80$c\,. A higher value would increase
the over saturation of QGP phase space. 
Despite a much greater expansion velocity,  the evolution
time at LHC is significantly longer compared to LHC, with the most central collisions taking up
to 30\% longer to reach the freeze-out hadronization temperature $T_f=0.14$--0.17 GeV.
This is due to the greater initial entropy and energy content that takes longer to dilute to 
the hadronization condition.

Finally, we need to evaluate the invariant strangeness production rate per unit 4-volume, 
 $A^{gg\to s\bar s}$ and $A^{q\bar q\to s\bar s}$,  see \req{qprod3a}.
These are obtained employing the running
strength of the QCD coupling,  with $\alpha_s(\mu=m_{Z^0})=0.118$  
 evolved  to the applicable energy domain $\mu$  using the two loops $\beta$-function. 

Strangeness production, a relatively soft process, can be 
described perturbatively for two reasons:\\
a) the reaction processes which change yield of strangeness 
can compete with the fast $v_\bot>0.5c$ expansion of   QGP only for 
$T>T_s=220\,{\rm MeV}\simeq 2m_s$, for lower temperatures the strange quark yields effectively
do not change. $T_s$ is the strange quark chemical freeze-out temperature in QGP. 
Using the relation $\mu=2\pi T$, this implies that all
strangeness yield evolution occurs for   $ \mu > 1.4 $\,GeV.\\
b) Because of the magnitude   $\alpha_s(\mu=m_{Z^0})=0.118$, one can 
run  $\alpha_s$ to the scale of interest, $\mu>1.2$GeV, 
and the value $\alpha_s/\pi\lessim 0.2$  allows perturbative QCD methods.\\

\begin{figure*}[t]
\vskip -0.5cm
\psfig{width=6.5cm,figure=\pathnow    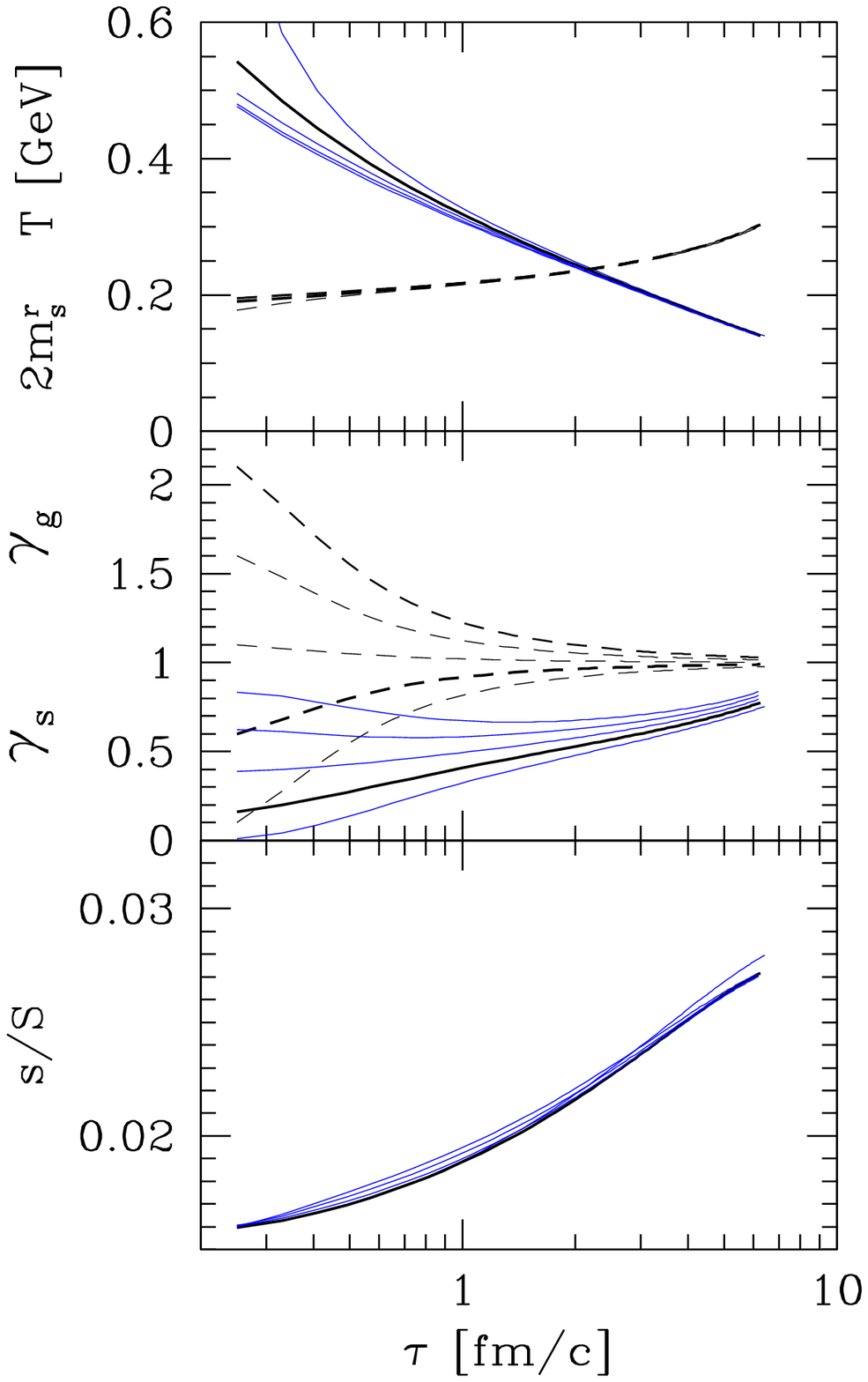  }\hspace*{-0.5cm}
\psfig{width=6.5cm,figure=\pathnow  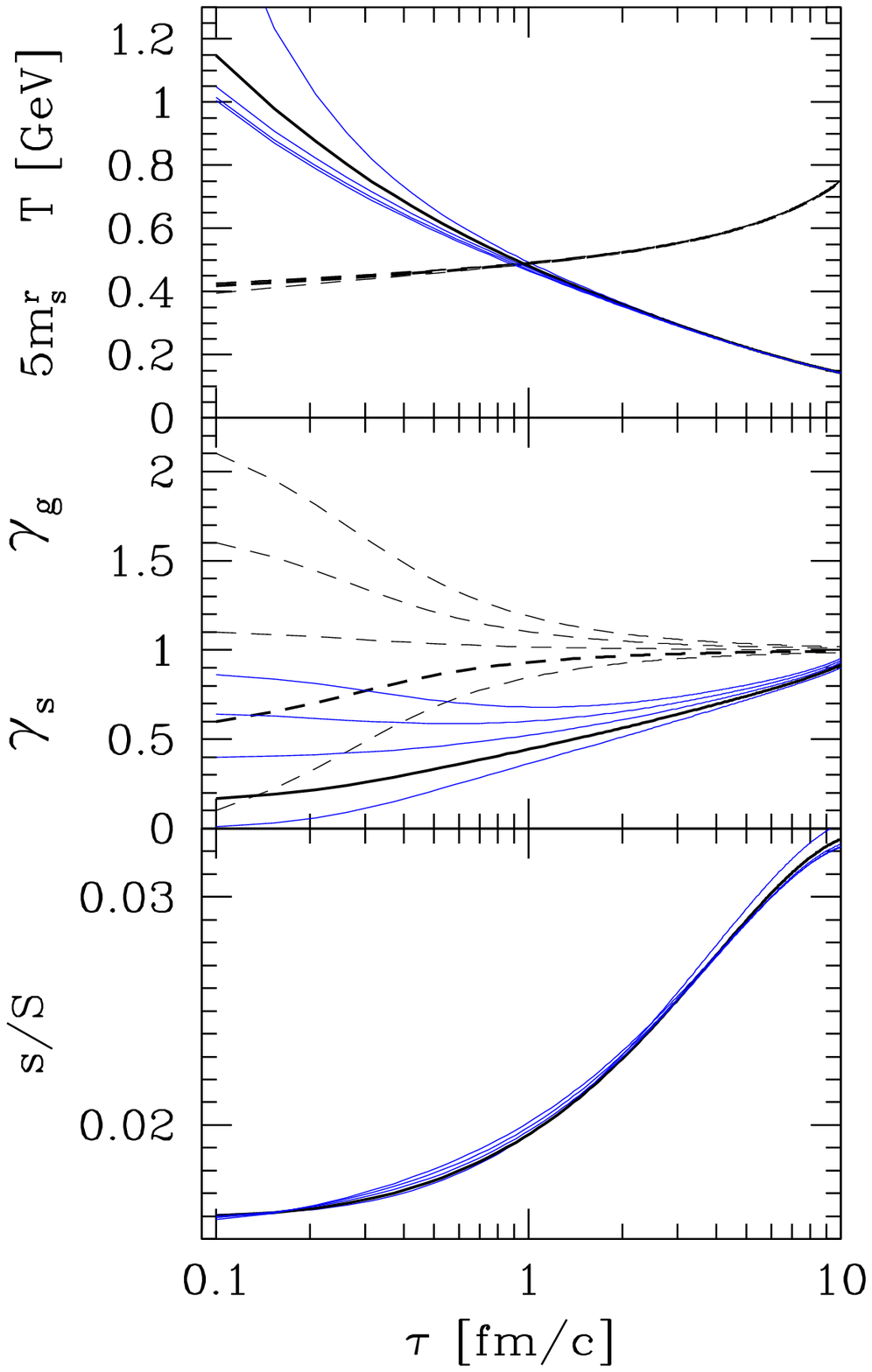   }
\vspace*{-0.6cm}
\caption{\label{Gluedep}
Strangeness production at RHIC (left) and 
LHC (right) for the case of  5\% most central collisions. 
 $s/S$ (bottom panel) and $\gamma_s$ (solid lines middle panel) 
evolution as function of proper time $\tau$. In this study,  widely 
varying initial gluon conditions ($\gamma_{\rm g}$ dashed middle panel)
  constrained to same entropy content, and thus, varying temperature $T$ (top panel),
is presented.
}
\end{figure*}

We also evolve  the central strange quark mass value  $m_s(\mu=2\,{\rm GeV})=100\pm25$ MeV.
 We use as the energy scale the CM-reaction energy 
$\mu\simeq  \sqrt{s}$  working in two loops. Some simplification is 
achieved by considering at temperature $T$ the QCD scale  $\mu\simeq 2\pi T$,
thus    $m_s(T)=m_s(\mu=2\pi T)$
with $m_s(T=318\,{\rm MeV})=0.1\,$GeV. The actual temperature  $T(t)$, and thus the   time 
dependent  values of the strange quark mass, are seen in the
top panel of figure~\ref{Gluedep}.

The figure \ref{Gluedep} illustrates in detail how a variation of the unknown
 initial chemical equilibration conditions impacts
the  final strangeness yield results. Since the entropy seen at the end of 
the reaction is already present at an early time, it is fixed as function of time
and is the same for the different initial state  scenarios explored.
$T(\tau_0)$ is obtained (top panel) for 
 a wide range of assumed initial gluon (and quark) occupancy
 $0.1<\gamma_{\rm g}(\tau_0)<2.1 $, in step of 0.5, 
shown in the middle panel
by dashed lines.

Results for   strangeness production at RHIC (left) and LHC (right) as function of 
proper time $\tau$ are shown in middle and bottom panels;
the solid lines in middle panel show the 
resulting $\gamma_s$, and in the bottom panel $s/S$. 
Two results emerge: the value of $\gamma_s$ at QGP hadronization  
is nearly independent of the initial conditions; the evolution of   $s/S$
in time  is   a universal curve.  A large part of the  difference between 
RHIC and LHC arises because the expansion lasts  longer. 
At RHIC, the thermal strangeness production raises the 
value of $s/S$ from 0.016 to 0.028, and
at the LHC the thermal strangeness production raises $s/S$ from
 0.016 to 0.032.  This 20\% increase in relative strangeness yield has 
a highly significant impact on production of strange hadrons.

These results show that the observable $s/S$ is dependent on the
dynamics of dense matter evolution (and also QCD parameters $\alpha_s, m_s$,
 but not on the internal structure of the QGP, and in particular not on the degree of chemical 
equilibration). At a fixed
entropy  yield, a  plasma that has not yet reached chemical equilibrium is hotter.
This higher $T$   compensates   in strangeness production  the lack of 
chemical equilibration, \ie,   $\gamma_G<1$.  Entropy is practically constant since 
 the process of chemical equilibration
does not produce significant amount of entropy~\cite{Letessier:1993qa}, which is 
mostly produced in the process
of thermal equilibration. 

The final strangeness content expressed by $s/S$ 
depends on the initial entropy content,  $dS(\tau_0)/dy$. This value can change 
both due to change in energy and centrality of the reaction. 
A scenario study shows that  variation  of centrality leaves
us on the universal $s/S$ curve seen in \rf{Gluedep}, but the freeze-out temperature
is reached sooner for smaller volumes with smaller $dS/dy$, 
and thus, smaller $s/S$ is achieved. A similar
result is obtained when varying reaction energy. For this reason, we
present    in     \rf{sSdy}  the final  $s/S$ and $\gamma_s$  as function of $dS/dy$.
 These two results, $s/S$ and $\gamma_s$,
 will change independently  under variation of strange quark mass, with 
$\gamma_s$ remaining constant and $s/S$ changing. 
Considering extreme conditions for the volume 
evolution one can vary both $s/S$ and $\gamma_s$ by the same magnitude, 10--20\%.
We remind the reader that the $\gamma_s$ discussed in this section is for the QGP
phase while the  $\gamma_s$ discussed in hadronization analysis is for HG phase
and is generally 2.5 times greater. 
 
\begin{figure*}[t] 
\vskip -0.5cm
\centerline{\psfig{width=7.9cm,figure=\pathnow   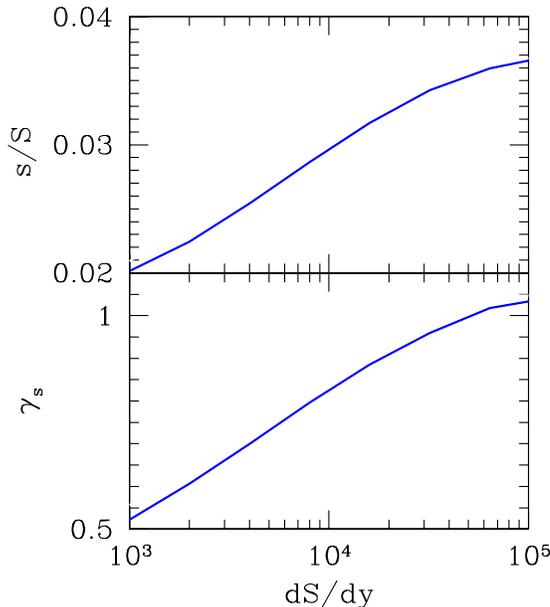   }
}
\vspace*{-0.6cm}
\caption{\label{sSdy}
 $s/S$ (top panel) and $\gamma_s$ (bottom panel) 
as function of   $dS/dy$.  
}
\end{figure*}

\section{Strange Hadron Resonances}\label{StrRes}
Theoretical models of hadron production need to account in their study of 
particle spectra or/and  hadron  yields for the contribution 
 originating in resonance decays. In the statistical hadronization,
 the production of hadron resonances is abundant. 
 Within a particle `family',  particle yields with same valance quark
content are in relation to each other well described by integrals 
of relativistic phase space, eventually corrected for 
rescattering~\cite{Rafelski:2001hp}. The relative yields of, \eg,
$K^*(\bar s q)$ and $K(\bar s q)$,  are  controlled 
by the particle masses $m_i$,  statistical weights (degeneracy) $g_i$ and the 
hadronization temperature $T$. In the Boltzmann limit,
one has (star denotes the resonance): 
\begin{equation}\label{RRes}
{N^*\over N}= \frac{g^*m^{*\,2}K_2(m^*/T)}{ g\,m^{2}K_2(m/T)}
    \frac{|\overline{{\cal M}^*}|^2}{|\overline{{\cal M}}|^2}\,.
\end{equation}

In the statistical hadronization model, the assumption made is that 
the matrix element ${\cal M}(E)$ is not  strongly energy dependent 
($E$ is energy of particle produced). Thus, as
shown above, the energy average factorizes. Moreover, one also expects that for most
strongly interacting particles,  the quantum strength 
saturates at its maximum, $|{\cal M}|^2\to 1$, $|{\cal M}^*|^2\to 1$,
in the energy range in which most particles are produced (but not necessarily
in some extreme corners of phase space where special situations prevail).

Resonance production is measured, in an experiment,  directly  by the decay product
invariant mass method. Presence of these particle, and the
fact that their 4-momenta add up to form invariant mass of the
parent  means that most if not all of the decay products do not
significantly rescatter. Except for special cases ($\phi$, $\Omega,\overline\Omega$)  
resonances contribute 50\% and more 
of the yield of each  hadron. When this is recognized in
an analysis of experimental data, one finds that all particles
are produced at the same temperature and that the yield temperature
(chemical freeze-out) and spectral temperature (thermal freeze-out) are
the same. This analysis was done for the maximal SPS energy~\cite{Torrieri:2000xi}, and the 
 maximal RHIC energy~\cite{Flor}. Claims to the contrary rely on analysis which 
rescatter the resonance decay products such that all particles
considered are forced to satisfy a thermal spectrum overlaid with flow from expansion
dynamics. This, then, produces for different particles different freeze-out 
conditions of temperature and velocity. Such analysis is of course wrong as it presumes tacitly
that not a single resonance should be observable by invariant mass method. 

Putting aside the global analysis of data, we can argue
qualitatively for single freeze-out,  comparing the 
baryon and antibaryon $m_\bot$ spectra  as measured 
very precisely by   WA97 and NA57 at SPS~\cite{Antinori:2000sb,Bruno:2004pe},
and STAR at RHIC~\cite{Adams:2003fy,Adams:2006ke}. We note that 
if hadron gas were to surround at any relevant stage the plasma phase, 
such a symmetry could not arise in  the presence of a baryon excess:
the  $\overline\Lambda, \overline\Xi$ annihilation  
 is strongly momentum dependent, this  
should deplete the low momentum yield. This   deforms  
the shape of the  antihyperon spectra, as compared to the 
spectra of hyperons. More  specifically,  $  \overline\Xi$
could annihilate on relatively abundant 
$N-\overline N,\Lambda-\overline \Lambda,\Sigma-\overline\Sigma$  
(rather than the much smaller $ \Xi -\overline\Xi$),
and yet the spectral $m_\bot$-distributions of  $  \overline\Xi$ 
are within a fraction of percent error, identical to  $ \Xi$. Absence 
of   spectral deformation   suggests  that $  \overline\Xi$  emerge directly and 
without significant rescattering.  

Our  working hypothesis is therefore
 that hadronization of the QGP deconfined phase formed
in high energy nuclear collisions at top SPS energy and 
RHIC energies is direct, fast (sudden) and 
occurs without significant sequel interactions.
In this circumstance, 
the measurement of the relative yield of hadron resonances 
is a sensitive test  of the statistical hadronization model (SHM) hypothesis
and lays the foundation  to the application of the SHM method in data analysis. 
One of the great successes of the chemical nonequilibrium  particle yield 
analysis is the agreement with the experimental
resonance yields.  These, in particular, are at 
RHIC K$^*(892),\Lambda(1520)$~\cite{Markert:2002rw,Markert:2005jv}.
Our relatively low hadronization (chemical freeze-out)  temperature 
makes this possible. More data is on the way and will keep us busy
in coming years~\cite{Adams:2006yu,Salur:2006jq}.

We believe that the systematic study of resonance production
can resolve the question about the reaction energy dependence of the 
freeze-out condition temperature. One could argue naively 
that this is simply the  temperature at the 
quark-hadron transformation seen in lattice computations.
This implies a smooth change with reaction energy, 
since temperature rises smoothly while baryon density
decreases. 
The maximum value of $T$ is reached at highest reaction energy.
However, since there is no sharp phase transition at small baryon density
prevailing at LHC, RHIC and perhaps top SPS energy,  there is 
in this hypothesis a lot of uncertainty:   the value of 
cross-over temperature varies depending on observables 
considered. Moreover,  among lattice  results there is considerable difference
in the range of interest for maximum $T$, with $T=195\pm10$ MeV~\cite{Cheng:2006qk},
and   $T=151\pm5$ MeV~\cite{Katz}, at zero baryon density. In recent years,
$T$  has varied within these limits~\cite{Karsch:2001vs,Fodor:2004nz,Karsch:2003jg} .

Our  view is that freeze-out temperature is, to a large extent, 
influenced by the dynamics of the rapidly expanding fireball~\cite{Vienna}:
a) When  the collective flow occurs at parton level,
the color charge flow, like a wind,   pushes  out the vacuum,
adding to thermal pressure a  dynamical component~\cite{suddenPRL,Csorgo:2002kt}. 
This  can, in general, lead to supercooling and a sudden breakup of 
the fireball. We find that this  can 
reduce the effective hadronization temperature 
by  20 MeV~\cite{suddenPRL}.\\
b)
We also must consider the chemical composition of the hadronizing QGP. It
is well known that the critical temperature is
dependent on the flavor content. Thus, the degree of chemical
equilibration of strange quark flavor  influences the 
exact location of the boundary, and the nature of the 
transition. Because the degree of 
equilibration in the QGP depends on the collision energy, as does 
the collective expansion velocity, we cannot   expect a 
simple hadronization scheme appropriate for the hadronization
of nearly adiabatically expanding Universe. 
We can be nearly sure that the chemical conditions matter and can
displace the transition temperature.

\begin{figure*} 
\centerline{\epsfig{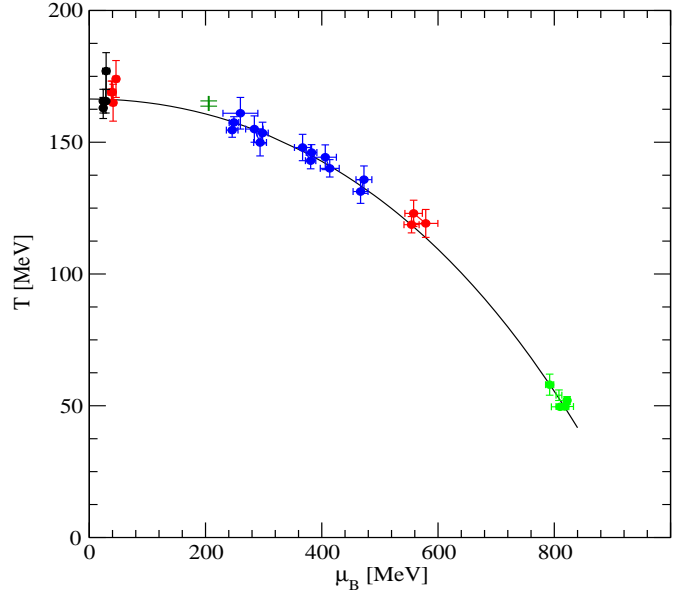}}
\caption{\label{CleyFig} 
Freeze-out boundary temperature  $T$ and baryo-chemical potential $\mu_{\rm B}$
on reaction energy following from  the chemical equilibrium hypothesis, after Ref.~ \cite{Cleymans:2006qe}.
The line guides the eye.}
\end{figure*}

The question of considerable importance is how we can find agreement on what the 
temperature of hadronization is, and be able to trace in the phase diagram 
the hadronization boundary, in a way that leads to general agreement. 
The disagreement is considerable.  In  figure \ref{CleyFig}, we see 
the situation proposed  for the hadron
system considered to be in absolute 
chemical equilibrium~\cite{Cleymans:2006qe}.
 
 At highest heavy ion reaction energy,  
one obtains chemical freeze-out temperature   $T \sim  170$ MeV. 
Values  as low as $T \sim 50$ MeV are reported at lowest reaction energies available.
As the collision energy increases,  the freeze-out temperature 
increases and the baryonic density (here baryonic chemical potential $\mu_{\rm B}$)
decreases.  Such an  increase of freeze-out temperature with collision energy is 
expected on general grounds, since  with increasing reaction energy a 
greater fraction of the energy is carried by mesons created in the collision, 
rather than pre-existing baryons~\cite{Hagedorn:1980kb}.

There is no
sign of any structure, in  figure \ref{CleyFig}, when the reaction energy varies. The rather smooth 
hadronization curve makes it impossible to describe the experimental 
discontinuity in the ratio K$^+/\pi^+$ (see figure \ref{kpiratio}) and other related experimental 
features~\cite{Gaz}.
An effort was made to interpret this in terms of a shift from baryon to meson dominance~\cite{hornthermal}
of the hadron yields. However,  in the chemical equilibrium model, all  observables including 
K$^+/\pi^+$ remain a smooth function of reaction energy, in contrast to the
experimental results. 
  
\begin{figure*} 
\centerline{\epsfig{width=9.7cm,height=9cm,clip=,figure=\pathnow 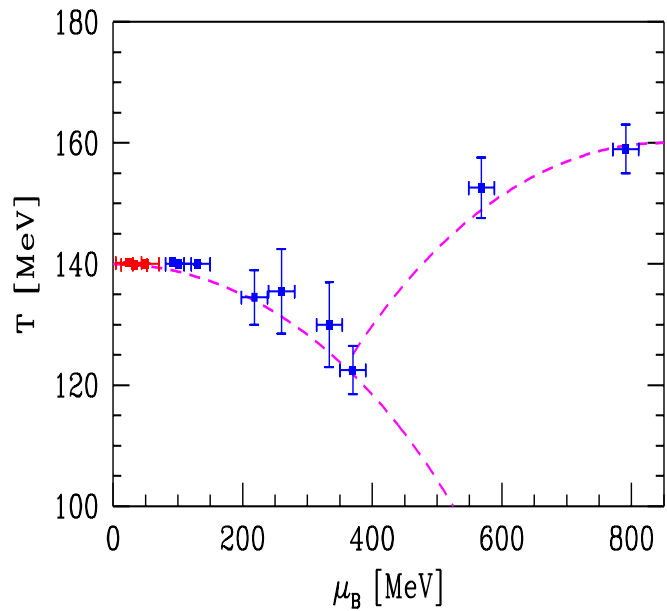}}
\vskip -0.4cm
\caption{\label{Ph} 
Dependence of freeze-out temperature  $T$ and baryo-chemical potential $\mu_{\rm B}$
on reaction energy in the non-equilibrium case~\cite{Letessier:2005qe}. The dashed lines
guide the eye.  }
\end{figure*}

The   systematic behavior of freeze-out condition in the 
$T$-$\mu_{\rm B}$ plane  allowing for $\gamma_q\ne 1$ 
is quite different~\cite{Letessier:2005qe}, see
figure \ref{Ph}. 
The two higher $T$ values  at right are 
for 20 $A$ GeV (lowest CERN-SPS energy) and  
(furthest to right) 11.6 $A$ GeV (highest BNL-AGS energy)
reactions. In these two cases the source of particles is   relatively large, dilute  and 
strongly chemically under-saturated, and  hotter than otherwise has been expected.
Such a system could be a conventional hadron gas fireball that 
had not the time to chemically equilibrate. Other options were   
 considered  such as a phase of constituent massive quarks. 

The higher temperature
and lower density imply that at the time of freeze-out the fireball is not 
expanding as fast as found at  higher reaction energies. Thus without 
doubt sudden hadronization does not apply here. 
The transition to the supercooled  regime occurs and the phase space changes
from chemically under-saturated ($\gamma_q<1$) to chemically over-saturated ($\gamma_q>1$)
where the two dashed  lines meet in figure~\ref{Ph}.
At heavy ion reaction energy below (\ie, to right in figure~ \ref{Ph}) of this point, hadrons freeze-out 
with meson number enhanced with reference 
to antibaryon yield. Above 30 $A$ GeV the antibaryon to meson yield is
enhanced. 

The relative resonance yield \req{RRes} 
is only dependent on the temperature of hadronization.
Hence the relative yield of resonances  will distinguish the two scenarios presented
in figures \ref{CleyFig} and \ref{Ph}~\cite{Torrieri:2006yb}. 
Torrieri \etal evaluated 
the   relative yield of resonances along the two hadronization
curves as function of reaction energy and have found that the 
behavior is practically opposite. Relative resonance yields in the equilibrium model
rise monotonically with energy of heavy ion reaction.  
The nonequilibrium model implies overall a reduction 
of $T$ with increasing heavy ion reaction energy, with a strong dip near to 
the `horn' energy. Moreover, and beyond the argument presented in~\cite{Torrieri:2006yb}, 
 the visible resonance yield could be negligible at lowest SPS energies due to rescattering 
of decay products in absence of   sudden hadronization.

\section{Outlook}\label{Outlook}
Rereading many experimental and theoretical works we have 
had impression that there is more experimental data than theoretical effort in
this very rich field. To our despair, much of the data analysis is scope-restricted and does
not seek overall consistency.  Authors of failed theoretical ideas do not come 
forward to withdraw. Their ideas live on, and are  resurrected.  
Today, we see more confusion in the field than there has ever been. Unfortunately,
this confusion is  welcome 
in some quarters, simply because there is such good strangeness data,
generating   a very clear message, overshadowing other observables.

At SPS and RHIC, we see a chemically equilibrated 
strangeness rich $s$-$q$-$g$-system. What we are not 100\% sure of
is, if this is every-day quark--gluon plasma, though 
there is plenty of evidence that it is. We have discussed
how the degree of chemical equilibration
in QGP varies, with system size, and reaction energy. The result follows  
from glue-based strangeness production. Entropy content seen in
particle multiplicity, is  for the most central and most energetic RHIC 
reactions just as QGP predicts ---   
the ratio $s/S$ measures directly the ratio of strange to all QGP degrees of
freedom. At the 10\% level, or better, it agrees  
with what one expects from  the chemically equilibrated, interacting quark--gluon fluid.  

Hadronization of the fireball of matter formed in heavy ion reactions leads to quite different 
spectra and yields of hadrons than we expect based on elementary $pp$ reactions. This
change in reaction mechanism favors, in particular, the production of multi strange baryons and antibaryons. 
The enhancement we see is what statistical recombination of quarks 
predicts, both as function of centrality and energy. 

Many years ago, when the first ideas how to look for quark-matter were born, it seemed
that the energy threshold to deconfinement could be low. Nucleons in
nuclei seemed to be just barely retaining their individual identities. Just a little push and squeeze 
with a few GeV beam could perhaps be sufficient 
to lead to deconfinement~\cite{abundance}.   This is where we are
today, after a long and dramatic excursion to the very high RHIC energies:
Strange antibaryon enhancement 
 suggests that at least down to 40 $A$ GeV we have  $s$-$q$-$g$-matter. 
The simplest of all possible observables, the K$^+/\pi^+$ ratio shows 
a threshold between 20 and 30 $A$ GeV projectile energy.  

Let us cross our  fingers that  RHIC can and will run at a few GeV per beam. The existent
detectors would in this environment produce very precise and have nearly
full coverage in phase space for the data on strange hadron production, including 
resonances. This would with certainty resolve any doubt about QGP, both its formation
and threshold as function of centrality and reaction energy. This will  
further lead to detailed understanding of the phases of QCD. This work
will complement the LHC based study of perturbative QGP.

\section*{Acknowledgments}
Supported  by a grant from the U.S. Department of
Energy,  DE-FG03-95ER40937\,. Laboratoire de Physique Th\'eorique 
et Hautes Energies, \linebreak LPTHE, at  University Paris 6 and 7 is supported 
by CNRS as Unit\'e Mixte de Recherche, UMR7589.


\begin{thebibliography}{10}



\bibitem{KMRbook}
J. Kapusta, B. M\"uller and J. Rafelski,
{\it quark--gluon Plasma: Theoretical Foundations}
(An annotated reprint collection) Elsevier (Amsterdam 2003).

 


\bibitem{share}
  G.~Torrieri, S.~Steinke, W.~Broniowski, W.~Florkowski, J.~Letessier and J.~Rafelski,
  ``SHARE: Statistical hadronization with resonances'',
  Comput.\ Phys.\ Commun.\  {\bf 167}, 229 (2005)
  [arXiv:nucl-th/0404083];\\
  G.~Torrieri, S.~Jeon, J.~Letessier and J.~Rafelski,
  ``SHAREv2: Fluctuations and a comprehensive treatment of decay feed-down'',
   Comput.\ Phys.\ Commun.\  in press [arXiv:nucl-th/0603026].





\bibitem{Rafelski:1980rk}
  J.~Rafelski and R.~Hagedorn,
  ``From Hadron Gas To Quark Matter. 2'',
CERN-TH-2969, in Satz, H. ed. {\it Statistical mechanics of quarks and hadrons},
North-Holland,  ISBN 0444862277, (Amsterdam, 1981) pp 253-272.
  
 
\bibitem{abundance}
  J.~Rafelski,
  ``Extreme States of Nuclear Matter'', In proceedings of {\it Future Relativistic Heavy Ion Experiments 1980},
R. Bock and R. Stock, Eds.,  Published by GSI, Darmstadt 1981, GSI Report 1981-6, pp 282-324;
scan available at:\\
http://www.physics.arizona.edu/\~{ }rafelski/PS/81GSIPaperJR.pdf

 
\bibitem{actab} 
  J.~Rafelski, J.~Letessier and A.~Tounsi,
  ``Strange Particles from Dense Hadronic Matter'',
  Acta Phys.\ Polon.\ B {\bf 27}, 1037 (1996)
  [arXiv:nucl-th/0209080  (late submission to make paper accessible)].


\bibitem{Greiner:1987tg}
  C.~Greiner, P.~Koch and H.~Stoecker,
``Separation of Strangeness from Antistrangeness in the Phase Transition from 
Quark to Hadron Matter: Possible Formation of Strange Quark Matter in Heavy Ion Collisions'',
  Phys.\ Rev.\ Lett.\  {\bf 58} (1987) 1825.




\bibitem{Alt:2006dk}
  C.~Alt {\it et al.}  [NA49 Collaboration],
   ``Energy and centrality dependence of anti-p and p production and the
   $\overline\Lambda/\bar p$   ratio in Pb-Pb collisions between 20 $A$ GeV and
  158 $A$ GeV'',
  Phys.\ Rev.\ C {\bf 73}, 044910 (2006)
  [arXiv:nucl-ex/0512033],   and private communication by M. Mitrovski.
  
  
 
\bibitem{Antinori:2006ij}
  F.~Antinori {\it et al.}  [NA57 Collaboration],
   ``Enhancement of hyperon production at central rapidity in 158 $A$ GeV/c Pb-Pb
  collisions'',
  J.\ Phys.\ G {\bf 32}, 427 (2006)
  [arXiv:nucl-ex/0601021].


\bibitem{Koch:1986ud}
  P.~Koch, B.~Muller and J.~Rafelski,
  ``Strangeness In Relativistic Heavy Ion Collisions'',
  Phys.\ Rept.\  {\bf 142}, 167 (1986).

\bibitem{Antinori:2004ee}
  F.~Antinori {\it et al.}  [NA57 Collaboration],
   ``Energy dependence of hyperon production in nucleus nucleus collisions  at
  SPS'',
  Phys.\ Lett.\ B {\bf 595}, 68 (2004)
  [arXiv:nucl-ex/0403022].


\bibitem{Antinori:2006rk}
  F.~Antinori {\it et al.},
  ``NA57 results'',
  AIP Conf.\ Proc.\  {\bf 828} (2006) 333.


\bibitem{Rafelski:2003ju}
  J.~Rafelski and J.~Letessier,
   ``Strangeness and statistical hadronization: How to study quark--gluon
  plasma'',
  Acta Phys.\ Polon.\ B {\bf 34}, 5791 (2003)
  [arXiv:hep-ph/0309030].

\bibitem{Rafelski:1999xu}
  J.~Rafelski and J.~Letessier,
  ``Diagnosis of QGP with strange hadrons'',
  Acta Phys.\ Polon.\ B {\bf 30}, 3559 (1999)
  [arXiv:hep-ph/9910300].


\bibitem{Caines:2006if}
  H.~Caines,
  ``Is soft physics entropy driven?'',
  arXiv:nucl-ex/0609004.



\bibitem{Mitrovski:2006js}
  M.~K.~Mitrovski {\it et al.}  [NA49 Collaboration],
  ``Strangeness production at SPS energies'',
  arXiv:nucl-ex/0606004,  J. Phys. G in press,
 


\bibitem{Letessier:2005kc}
  J.~Letessier and J.~Rafelski,
  ``Centrality dependence of strangeness and (anti)hyperon production at  RHIC'',
  Phys.\ Rev.\ C {\bf 73}, 014902 (2006)
  [arXiv:nucl-th/0506044].



\bibitem{Caines:2004ej}
  H.~Caines  [STAR Collaboration],
  ``Volume effects on strangeness production'',
  J.\ Phys.\ G {\bf 31}, S1057 (2005).
 


 \bibitem{phenixyield} 
S.~S.~Adler {\it et al.}  [PHENIX Collaboration],
``Identified charged particle spectra and yields in Au-Au collisions at
s(NN)**(1/2) = 200-GeV'',
Phys.\ Rev.\ C {\bf 69}, 034909 (2004)
[arXiv:nucl-ex/0307022].
 

\bibitem{Rafelski:2004dp}
  J.~Rafelski, J.~Letessier and G.~Torrieri,
  ``Centrality dependence of bulk fireball properties at RHIC'',
  Phys.\ Rev.\ C {\bf 72}, 024905 (2005)
  [arXiv:nucl-th/0412072].
 
\bibitem{Redlich:2001kb}
  K.~Redlich and A.~Tounsi,
  ``Strangeness enhancement and energy dependence in heavy ion collisions'',
  Eur.\ Phys.\ J.\ C {\bf 24}, 589 (2002)
  [arXiv:hep-ph/0111261].


  
\bibitem{star-b1} C. Adler {\it et al.}  [STAR Collaboration], 
``Midrapidity $\Lambda$  and  $\overline\Lambda$  production in Au-Au Collisions at $\sqrt{s_{\rm NN}}=130$ GeV''
Phys. Rev. Lett. {\bf 89}, 092301 (2002).

\bibitem{star-b2} 
  J.~Adams {\it et al.}  [STAR Collaboration],
   ``Identified particle distributions in $p p$ and Au-Au collisions at $\sqrt{s_{\rm NN}}=200$ GeV''
  Phys.\ Rev.\ Lett.\  {\bf 92}, 112301 (2004)
  [arXiv:nucl-ex/0310004].

\bibitem{star-k} J. Adams {\it et al.}   [STAR Collaboration], 
 ``Kaon production and kaon to pion ratio in Au-Au collisions at $\sqrt{s_{\rm NN}}=130$ GeV''
Phys. Lett. B{\bf 595}, 143 (2004).


\bibitem{Duke}  
  R.~J.~Fries, B.~Muller, C.~Nonaka and S.~A.~Bass,
   ``Hadronization in heavy ion collisions: Recombination and fragmentation  of
  partons'', and ``Hadron production in heavy ion collisions: Fragmentation and  recombination
  from a dense parton phase'',
  Phys.\ Rev.\ Lett.\  {\bf 90}, 202303 (2003) and Phys.\ Rev.\ C {\bf 68}, 044902 (2003)
  [arXiv:nucl-th/0301087,nucl-th/0306027].

\bibitem{Hwa}  
  R.~C.~Hwa and C.~B.~Yang,
  ``Scaling behavior at high $p_\bot$  and the $p/\pi$ ratio'',
  Phys.\ Rev.\ C {\bf 67}, 034902 (2003)
  [arXiv:nucl-th/0211010].


\bibitem{Letessier:2005qe}
  J.~Letessier and J.~Rafelski,
   ``Hadron production and phase changes in relativistic heavy ion
  collisions'',
  arXiv:nucl-th/0504028, submitted to PRC.


\bibitem{Gaz}
M.~Gazdzicki {\it et al.}  [NA49 Collaboration],
``Report from NA49'',
J.\ Phys.\ G {\bf 30}, S701 (2004)
[arXiv:nucl-ex/0403023], and   commented compilation of NA49 
results, private communication. 


 
\bibitem{Lee03}
M.~vanLeeuwen,
\newblock Compilation of {NA49} results as function of collision energy.
\newblock Private communication (2003).

\bibitem{Glendenning:1984ta}
  N.~K.~Glendenning and J.~Rafelski,
  ``Kaons and Quark Gluon Plasma'',
  Phys.\ Rev.\ C {\bf 31}, 823 (1985).




\bibitem{haibin2200}
H.~B.~Zhang  [STAR Collaboration],
``$\Delta$,  K* and $\rho$ resonance production and their probing of freeze-out
dynamics at RHIC'',, poster presentation at QM2004, Oakland, January 2004
[arXiv:nucl-ex/0403010];\\
J.~Adams  [STAR Collaboration],
``K$^*(892)$ resonance production in Au-Au and $pp$ collisions at $\sqrt{s_{\rm NN} }= 200$
GeV at STAR'',
 Phys.\ Rev.\ C {\bf 71}, 064902 (2005)
[arXiv:nucl-ex/0412019].

\bibitem{phiyld}
J.~Adams {\it et al.}  [STAR Collaboration],
``Phi meson production in Au-Au and$pp$ collisions at $\sqrt{s_{\rm NN}}=200$ GeV'',  
 Phys. Lett. B {\bf 612},   181 (2005)
[arXiv:nucl-ex/0406003].

\bibitem{Letessier:2004cs}
J.~Letessier, J.~Rafelski and G.~Torrieri,
``Deconfinement energy threshold: Analysis of hadron yields at 11.6-A-GeV'',
arXiv:nucl-th/0411047.
 


\bibitem{Letessier:2006wn}
  J.~Letessier and J.~Rafelski,
  ``Strangeness chemical equilibration in QGP at RHIC and LHC'',
  arXiv:nucl-th/0602047.



\bibitem{Chin:1979yb}
  S.~A.~Chin and A.~K.~Kerman,
  ``Possible longlived hyperstrange multi-quark droplets'',
  Phys.\ Rev.\ Lett.\  {\bf 43}, 1292 (1979).

\bibitem{Letessier:1993qa}
  J.~Letessier, J.~Rafelski and A.~Tounsi,
  ``Gluon production, cooling and entropy in nuclear collisions'',
  Phys.\ Rev.\ C {\bf 50}, 406 (1994)
  [arXiv:hep-ph/9711346].



\bibitem{Rafelski:2001hp}
  J.~Rafelski, J.~Letessier and G.~Torrieri,
  ``Strange hadrons and their resonances: A diagnostic tool of QGP  freeze-out
  dynamics'',
  Phys.\ Rev.\ C {\bf 64}, 054907 (2001)
  [Erratum-ibid.\ C {\bf 65}, 069902 (2002)].

\bibitem{Torrieri:2000xi}
  G.~Torrieri and J.~Rafelski,
  ``Search for QGP and thermal freeze-out of strange hadrons'',
  New J.\ Phys.\  {\bf 3}, 12 (2001)
  [arXiv:hep-ph/0012102].


\bibitem{Flor}
  W.~Florkowski,
  ``Particle spectra and hydro-inspired models'',
  Nucl.\ Phys.\ A {\bf 774}, 179 (2006)
  [arXiv:nucl-th/0509039];\\
  W.~Broniowski and W.~Florkowski,
  ``Explanation of the RHIC p(T)-spectra in a thermal model with expansion'', and 
  ``Strange particle production at RHIC in a single-freeze-out model'',
  Phys.\ Rev.\ Lett.\  {\bf 87}, 272302 (2001) and  Phys.\ Rev.\ C {\bf 65}, 064905 (2002)
  [arXiv:nucl-th/0106050 and arXiv:nucl-th/0112043].


\bibitem{Antinori:2000sb}
  F.~Antinori {\it et al.}  [WA97 Collaboration],
   ``Transverse mass spectra of strange and multi-strange particles in Pb Pb
  collisions at 158-A-GeV/c'',
  Eur.\ Phys.\ J.\ C {\bf 14}, 633 (2000).


\bibitem{Bruno:2004pe}
  G.~E.~Bruno  [NA57 Collaboration],
   ``Blast-wave analysis of strange particle $m_\bot$ spectra in Pb-Pb  collisions
  at the SPS'',
  J.\ Phys.\ G {\bf 31}, S127 (2005)
  [arXiv:nucl-ex/0410014].


\bibitem{Adams:2003fy}
  J.~Adams {\it et al.}  [STAR Collaboration],
   ``Multi-strange baryon production in Au Au collisions at $\sqrt{s_{NN}}  = 130$ GeV'',
  Phys.\ Rev.\ Lett.\  {\bf 92}, 182301 (2004)
  [arXiv:nucl-ex/0307024].


\bibitem{Adams:2006ke}
  J.~Adams {\it et al.}  [STAR Collaboration],
   ``Scaling properties of hyperon production in Au-Au collisions at
  $\sqrt{s_{NN}}  = 200$ GeV'',
  arXiv:nucl-ex/0606014.



\bibitem{Markert:2002rw}
  C.~Markert, G.~Torrieri and J.~Rafelski,
  ``Strange hadron resonances: Freeze-out probes in heavy-ion collisions'',
  arXiv:hep-ph/0206260, in {\it New states of matter in hadronic interactions},
  AIP Conference Proceedings {\bf 631}, pp 533-552. 

\bibitem{Markert:2005jv}
  C.~Markert,
  ``What do we learn from resonance production in heavy ion collisions?'',
  J.\ Phys.\ G {\bf 31}, S169 (2005)
  [arXiv:nucl-ex/0503013].


\bibitem{Adams:2006yu}
  J.~Adams {\it et al.}  [STAR Collaboration],
   ``Strange baryon resonance production in $\sqrt{s_{\rm NN}} = 200$ GeV $pp$ and  Au-Au collisions'',
  arXiv:nucl-ex/0604019, Phys. Rev. Lett. in press.

\bibitem{Salur:2006jq}
  S.~Salur,
  ``Baryonic resonance studies with STAR'',
  [arXiv:nucl-ex/0606002], J. Phys. G in press.






\bibitem{Cheng:2006qk}
  M.~Cheng {\it et al.},
  ``The transition temperature in QCD'',
  [arXiv:hep-lat/0608013].

\bibitem{Katz}
Y. Aoki, Z. Fodor, S.D. Katz, K.K. Szabo
``The QCD transition temperature: results with physical masses in the continuum limit'',
  [arXiv:hep-lat/0609068].



\bibitem{Karsch:2001vs}
F.~Karsch,
``Lattice results on QCD thermodynamics'',
Nucl.\ Phys.\ A {\bf 698}, 199 (2002)
[arXiv:hep-ph/0103314].
 
\bibitem{Fodor:2004nz}
Z.~Fodor and S.~D.~Katz,
``Critical point of QCD at finite T and mu, lattice results for physical quark
masses'',
JHEP {\bf 0404}, 050 (2004)
[arXiv:hep-lat/0402006].




\bibitem{Karsch:2003jg}
  F.~Karsch and E.~Laermann,
  ``Thermodynamics and in-medium hadron properties from lattice QCD'',
  [arXiv:hep-lat/0305025]. 
In R.C. Hwa,  et al.: Quark gluon plasma III (2004) pp 1-59\ (World
Scientific, Singapore).
 

 

\bibitem{Vienna}
J. Rafelski   and J. Letessier
``Hadronization of expanding QGP'', 
  Eur.\ Phys.\ J.\ A {\bf 29} 107  (2006)
  [arXiv: nucl-th/0511016].

\bibitem{suddenPRL}
J.~Rafelski and J.~Letessier,
``Sudden hadronization in relativistic nuclear collisions'',
Phys.\ Rev.\ Lett.\  {\bf 85}, 4695 (2000)
[arXiv:hep-ph/0006200].

\bibitem{Csorgo:2002kt}
  T.~Csorgo and J.~Zimanyi,
  ``Inflation of fireballs, the gluon wind and the homogeneity of the HBT
  radii at RHIC'',
  Heavy Ion Phys.\  {\bf 17}, 281 (2003)
  [arXiv:nucl-th/0206051].
 
 


\bibitem{Cleymans:2006qe}
  J.~Cleymans, H.~Oeschler, K.~Redlich and S.~Wheaton,
  ``Status of chemical freeze-out'',
  [arXiv:hep-ph/0607164].


\bibitem{Hagedorn:1980kb}
  R.~Hagedorn and J.~Rafelski,
  ``Hot hadronic matter and nuclear collisions'',
  Phys.\ Lett.\ B {\bf 97}, 136 (1980).


\bibitem{hornthermal} 
  J.~Cleymans, H.~Oeschler, K.~Redlich and S.~Wheaton, 
  ``The horn and the thermal model'', 
  [arXiv:hep-ph/0504065]. 


\bibitem{Torrieri:2006yb}
  G.~Torrieri and J.~Rafelski,
  ``Hadron resonances and phase threshold in heavy ion collisions'',
  [arXiv:nucl-th/0608061].

\end{thebibliography}


\end{document}